\documentclass{article} % For LaTeX2e
\usepackage{iclr2027_conference,times}

\usepackage{amsmath,amsfonts,bm}

\def\eqref#1{equation~\ref{#1}}
\def\1{\bm{1}}

\DeclareMathAlphabet{\mathsfit}{\encodingdefault}{\sfdefault}{m}{sl}
\SetMathAlphabet{\mathsfit}{bold}{\encodingdefault}{\sfdefault}{bx}{n}

\usepackage{hyperref}
\usepackage{url}
\usepackage{graphicx}

\usepackage{multicol}
\usepackage{multirow}
\usepackage{tabularx}
\usepackage{array}
\usepackage{makecell}
\usepackage{wrapfig}
\usepackage[table]{xcolor}
\usepackage{xcolor}
\usepackage{pifont}
\usepackage{enumitem}
\usepackage{amsmath}
\usepackage{amssymb}

\usepackage{booktabs}

\newcolumntype{L}{>{\raggedright\arraybackslash}X}
\newcolumntype{C}{>{\centering\arraybackslash}X}
\newcolumntype{R}{>{\raggedleft\arraybackslash}X}

\newcommand{\demph}[1]{\textcolor{gray!100}{#1}}
\newcommand{\cmark}{\textcolor{green!60!black}{\ding{51}}}
\newcommand{\xmark}{\textcolor{red!70!black}{\ding{55}}}

\title{V2SA-Bench: A Source-Aware Benchmark for Video-to-Spatial Audio Generation}

\author{Saksham Singh Kushwaha$^{1}$\thanks{Work done during an internship at Sony Group Corporation.} \quad Kazuki Shimada$^2$ \quad Koichi Saito$^2$ \\
 \textbf{Takashi Shibuya}$^2$ \quad \textbf{Yapeng Tian}$^1$ \quad \textbf{Yuki Mitsufuji}$^{2}$ \\ %[2pt]
$^{1}$The University of Texas at Dallas \quad 
$^{2}$Sony Group Corporation \\ [4pt]
\ Project page: 
\href{https://sakshamsingh1.github.io/v2sa-bench_webpage/}
{\textcolor{blue!50!black}{\texttt{https://sakshamsingh1.github.io/v2sa-bench\_webpage/}}}
}

\iclrfinalcopy 

\begin{document}

\maketitle

\begin{abstract}
Rapid growth in augmented and virtual reality applications has increased the focus on video-to-spatial audio generation, which conveys immersive cues about objects and their positions in space. However, evaluation efforts remain mono-centric and operate at the global-scene level rather than on individual sources. We introduce V2SA-Bench, a source-aware benchmark for video-to-spatial audio models, built on a human-annotated dataset of 1,274 videos across perspective and 360° videos, focusing on stereo and binaural audio and extending to first-order ambisonics. We define five evaluation dimensions, i.e., semantics, temporal, spatial, acoustics, and perceptual quality, with 12 fine-grained, reference-free, source-aware metrics that we show align well with human judgment. Across 13 baselines, we find that models are strongly center-biased, that spatial alignment is the largest bottleneck, and that performance drops from single to multiple sources. We further find that verifying source presence is essential and that higher perceptual quality does not imply better audio-visual alignment. We hope V2SA-Bench and its interpretable, source-aware metrics offer actionable directions for improving spatial audio generation. We will release our code and benchmark to facilitate future research.
\end{abstract} 
\section{Introduction}

\begin{wrapfigure}{r}{0.40\textwidth}
    \centering
    \vspace{-5mm}
    \includegraphics[width=\linewidth]{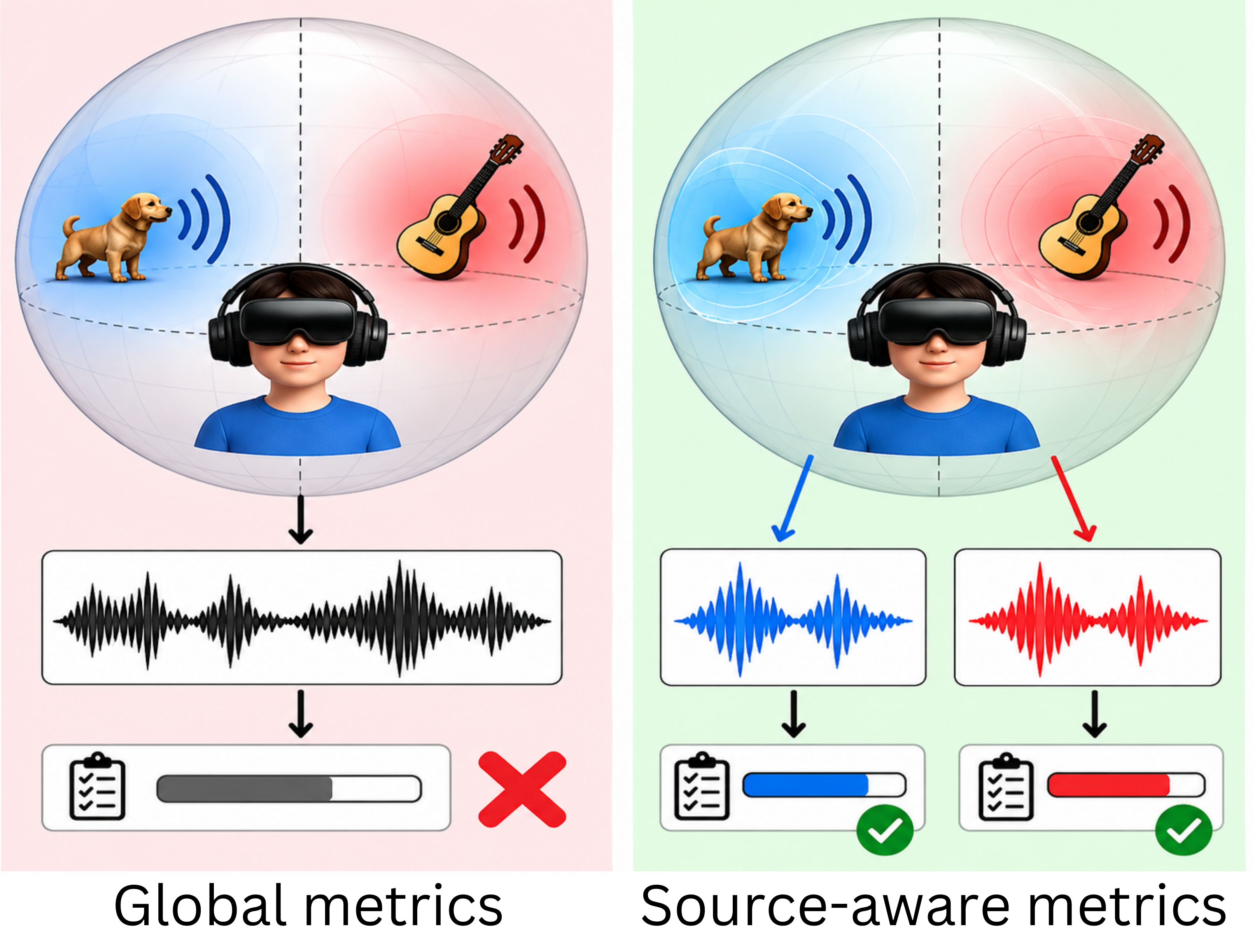}
    \vspace{-7mm}
    \caption{Global metrics overlook source-level mismatches, while our V2SA metrics explicitly evaluate the alignment of each active sound source.}
    \vspace{-2mm}
    \label{fig:teaser}
\end{wrapfigure}

The rapid growth of virtual reality (VR), augmented reality (AR), and metaverse applications, together with teleconferencing and film dubbing, has driven a push toward more immersive spatial audio. Unlike mono audio, spatial audio conveys a rich set of perceptual cues, i.e., directionality, distance, source location, room acoustics, and an overall sense of immersion. Manually authoring such spatial audio for these applications is labor-intensive and requires expert skills, motivating automatic generation directly from visual content. In particular, video-to-spatial audio generation is a natural fit, since the video itself provides strong cues about what sounds are present and where they originate. Building on rapid progress in text- and video-to-audio generation, a growing body of work now aims to synthesize spatial audio, i.e., stereo, binaural, and first-order ambisonics (FOA), directly from video, through both end-to-end models~\citep{tian2026audiox,kim2025visage,liu2025omniaudio} and multi-stage cascaded pipelines~\citep{gao20192}.

Yet, despite this progress, evaluation benchmarks lag behind. A first limitation is that they have remained overwhelmingly mono audio-centric~\citep{li2026benchmarking,dixit2026foleybench,zhang2026vidaudio}. While a few spatial-audio benchmarks exist, they support only a limited set of formats and remain narrow in scope.
As a result, existing benchmarks largely assess only what is generated while ignoring where it is localized, cannot verify the multi-channel structure of spatial formats, and fail to reveal spatial errors such as center bias.

A second limitation is that most existing metrics operate at the level of the global scene.
While a few recent works have begun to consider source-level cues, each is typically tied to a specific audio format or only a few dimensions. As a result, no benchmark unifies source-aware evaluation across dimensions and spatial formats. Yet humans perceive the world as a mixture of distinct sound sources, such as a fan whirring overhead while footsteps sound from below. This is precisely how professional object-based audio formats such as Dolby Atmos treat sound: as a set of sources, each with its own direction, semantics, and temporal behavior. Evaluation metrics should therefore mirror this perception and compute source-wise measures: given a piano and a guitar in a scene, we should be able to assess the generated audio for each source independently.

A third limitation lies in current evaluation metrics. Common measures such as Fréchet Audio Distance (FAD) \citep{gui2024fadtk} and KL divergence summarize distribution-level similarity but provide limited diagnostic insight into why a generated audio sample is incorrect or how a model should improve. Other metrics rely on reference audio, which may be unavailable for open-ended generation. We instead focus on objective, reference-free metrics that explicitly measure audio-visual alignment and provide interpretable feedback. Our design follows a simple intuition: \textit{if a human watches a muted video, what should they expect it to sound like?} To make this evaluation well-defined, we curate clips with clearly visible, active sources and unambiguous expected audio.

In summary, current audio evaluation suffers from three shortcomings: (1) it is overwhelmingly mono-centric and overlooks essential spatial cues; (2) it evaluates global scenes rather than individual sources; and (3) it often relies on abstract, distribution-based metrics or reference-dependent measures that offer limited interpretable, source-level feedback to future model developers.

To address these shortcomings, we introduce V2SA-Bench, a benchmark for video-to-spatial audio generation. First, to move beyond mono-centric evaluation, we cover two video and three spatial audio formats, focusing on stereo and binaural audio and extending to FOA, over a human-annotated corpus of 1,274 videos (767 perspective and 507 360-degree videos), each manually verified to contain active sound sources and annotated with their corresponding masks. Second, to move beyond global-scene evaluation, and as illustrated in Figure~\ref{fig:teaser}, we introduce 12 fine-grained metrics spanning five dimensions: semantic, temporal, spatial, acoustic, and perceptual quality. Most of these metrics are source-aware, operating at the level of individual sound sources, while a scene-level soundscape metric captures the overall acoustic environment and an audio-only metric assesses perceptual quality. Third, to overcome the reliance on distributional or reference-based metrics, our measures are reference-free and human-interpretable. Additionally, to enable a fair and extensive comparison, we evaluate 13 baselines spanning both the end-to-end and multi-stage cascaded paradigms.

Through comprehensive experiments and analyses, we provide, to the best of our knowledge, the first source-wise, quantitative evidence that current video-to-spatial audio models are strongly center-biased and that spatial alignment remains the largest bottleneck. Multi-stage cascaded pipelines tend to outperform end-to-end models on most dimensions except perceptual quality, though this advantage stems from a high-quality mono first stage rather than the cascaded design itself. Performance also drops consistently from single to multiple sources, and higher perceptual quality does not imply better audio-visual alignment. Finally, we show that verifying source presence is essential and that our metrics align well with human preference.

%%%%%%%%%%%%%%%%%%%%%%%%%%%%%%%%%%%%%%%%%%%%%%%%%%%%%%%%%%%%%%%%%%%%
\begin{table*}[t]
\centering
\caption{Comparison with existing benchmarks for conditional audio generation. Prior benchmarks primarily focus on mono audio generation, whereas V2SA-Bench evaluates video-to-spatial audio generation across 12 dimensions using reference-free metrics. Here, V = video, T = text, A = audio, PoV = perspective video, and FOA = first-order ambisonics.
}
\label{tab:benchmark_comparison}
\resizebox{\textwidth}{!}{
\begin{tabular}{lccccccccc}
\toprule
\textbf{Benchmark} &
\makecell{\textbf{Input}\\\textbf{Modality}} &
\makecell{\textbf{Output}\\\textbf{Modality}} &
\makecell{\textbf{Bench}\\\textbf{Size}} &
\makecell{\textbf{\# Eval}\\\textbf{Dimensions}} &
\makecell{\textbf{Source}\\\textbf{Aware}} &
\makecell{\textbf{All Reference}\\\textbf{Free metrics}} &
\makecell{\textbf{Spatial}\\\textbf{Metrics}} &
\makecell{\textbf{Acoustics}\\\textbf{Metrics}} &
\makecell{\textbf{Audio-visual}\\\textbf{Alignment}} \\
\midrule

FlatSounds\color{red}{$^1$}     & V (PoV) / T & A (Mono) & 185 & 9 & \xmark & \xmark & \xmark & \cmark & \cmark \\
FoleyBench\color{red}{$^2$}     & V (PoV) & A (Mono) & 5,000 & 6 & \xmark & \xmark & \xmark & \xmark & \cmark \\
VidAudio-Bench\color{red}{$^3$} & V (PoV) / T & A (Mono) & 1,634 & 13 & \xmark & \cmark & \xmark & \xmark & \cmark \\
TTA-Bench\color{red}{$^4$}      & T & A (Mono) & 2,999 & 7 & \xmark & \cmark & \xmark & \xmark & \xmark \\
AudioEval\color{red}{$^5$}      & T & A (Mono) & 4,200 & 5 & \xmark & \cmark & \xmark & \xmark & \xmark \\
T2A-EpicBench\color{red}{$^6$}  & T & A (Mono) & 100 & 5 & \xmark & \xmark & \xmark & \xmark & \xmark \\
VinTAGe-Bench\color{red}{$^7$}  & V (PoV) / T & A (Mono) & 636 & 9 & \xmark & \xmark & \xmark & \xmark & \cmark \\
MMAG\color{red}{$^8$}           & T & A (Mono) & 4,000 & 8 & \xmark & \xmark & \xmark & \xmark & \xmark \\
UniHAGen-Bench\color{red}{$^9$} & V (PoV) / T & A (Mono) & 1,003 & 5 & \xmark & \xmark & \xmark & \xmark & \cmark \\
Sphere360-Bench\color{red}{$^{10}$} & V (360°) & A (FOA) & 3,000 & 5 & \xmark & \xmark & \cmark & \xmark & \xmark \\

\midrule
\rowcolor{gray!15}
\textbf{V2SA-Bench} (Ours)  & V (PoV/360°) & A (Stereo/Binaural/FOA) & 1,274 & 12 & \cmark & \cmark & \cmark & \cmark & \cmark \\
\bottomrule
\multicolumn{10}{l}{\scriptsize \textcolor{red}{$^1$}~\citet{li2026benchmarking}, \textcolor{red}{$^2$}~\citet{dixit2026foleybench}, \textcolor{red}{$^3$}~\citet{zhang2026vidaudio}, \textcolor{red}{$^4$}~\citet{wang2026tta}, \textcolor{red}{$^5$}~\citet{wang2025audioeval}, \textcolor{red}{$^6$}~\citet{wang2025t2a}, \textcolor{red}{$^7$}~\citet{kushwaha2025vintage}, \textcolor{red}{$^8$}~\citet{zheng2026mmag}, \textcolor{red}{$^9$}~\citet{pian2026omnisonic}, \textcolor{red}{$^{10}$}~\citet{liu2025omniaudio}}
\end{tabular}
}
\vspace{-4mm}
\end{table*}
%%%%%%%%%%%%%%%%%%%%%%%%%%%%%%%%%%%%%%%%%%%%%%%%%%%%%%%%%%f

Our main contributions are summarized as follows:
\begin{itemize}
    \item \textbf{V2SA-Bench}. A comprehensive video-to-spatial audio generation benchmark spanning two video formats and three spatial-audio formats, comprising a high-quality dataset and source-aware metrics described below.
    \item \textbf{High-quality dataset}. A manually verified and annotated dataset with active sound sources and their corresponding masks.
    \item \textbf{Source-aware metrics}. Reference-free, human-aligned metrics across five dimensions and 12 fine-grained categories, most of which operate at the level of individual sound sources. Unlike prior source-level or spatial-audio metrics that are tied to a single audio format or a few dimensions, our metrics unify source-aware evaluation across semantic, temporal, spatial, and acoustic and across three spatial formats, with a semantic-gating design (validated in Sec.~\ref{sec:additional_analysis}) that prevents misleading scores from absent or wrong sources.
    \item \textbf{Valuable insights}. An extensive evaluation of 13 baselines providing, to the best of our knowledge, the first source-wise, quantitative characterization that current models are strongly center-biased with spatial alignment as the key bottleneck, and that our metrics align well with human judgment.
\end{itemize}
\section{Related works}

\noindent \textbf{Spatial audio generation}.
Early spatial audio was synthesized using signal-processing techniques such as room acoustic simulation~\citep{scheibler2018pyroomacoustics}, while subsequent deep-learning methods primarily spatialized existing mono audio~\citep{gao20192}. More recently, end-to-end models generate spatial audio directly from video, including video-to-stereo or binaural methods~\citep{tian2026audiox,liu2026thinksound,karchkhadze2026stereofoley,zhang2025visaudio,zhao2026foleyspace} and video-to-FOA methods~\citep{kim2025visage,liu2025omniaudio,lei2026towards}.
Cascaded pipelines instead first generate mono audio and then spatialize it using estimated source locations~\citep{gao20192,dagli2025see}.
Our benchmark evaluates such methods whose checkpoints are publicly available. These works also introduce spatial-audio datasets such as FAIR-Play~\citep{gao20192}, Sphere360~\citep{liu2025omniaudio}, and YT-Ambigen~\citep{kim2025visage}; we collect these datasets and manually verify each sample to build the V2SA-Bench dataset.

\noindent \textbf{Evaluation benchmarks}.
The rapid emergence of generative models has made benchmarks essential for measuring and driving progress. Comprehensive benchmarks exist for image and video generation~\citep{huang2023t2i,sun2025t2v,huang2024vbench}, and this maturity has recently extended to mono audio generation\citep{li2026benchmarking,dixit2026foleybench,zhang2026vidaudio}. However, these audio benchmarks operate on the global scene and are mono-centric, and thus cannot assess the spatial and source-wise cues central to spatial audio. Existing spatial metrics are also limited: the DOA-error metric of OmniAudio~\citep{liu2025omniaudio} (following ImmerseDiffusion~\citep{heydari2025immersediffusion}) and the saliency-based metric of ViSAGe~\citep{kim2025visage} both require ground-truth audio, while the bin-alignment score of StereoFoley~\citep{karchkhadze2026stereofoley}, which measures the alignment between object position and the audio energy center, cannot handle multiple simultaneous sources. Despite the growing importance of spatial audio for immersive applications such as virtual environments and world models~\citep{jin2026sonoworld}, no unified, source-aware benchmark exists.
We introduce V2SA-Bench, which evaluates video-to-spatial audio generation models across diverse audio categories with reference-free, human-aligned, and verifiable metrics. Table~\ref{tab:benchmark_comparison} provides a detailed comparison with existing conditional audio generation benchmarks in terms of scale, modality coverage, source-awareness, and evaluation capabilities.

\noindent \textbf{Source-aware audio generation}.
Recent work increasingly incorporates object and source-level conditioning into audio generation~\citep{karchkhadze2026stereofoley,li2025sounding,SS2A,viertola2026video,yoshida2026scene2sound,kushwaha2025diff}, using visual detection~\citep{SS2A}, segmentation masks~\citep{li2025sounding,viertola2026video}, or 3D source grounding~\citep{yoshida2026scene2sound}. Evaluation has similarly begun to consider source-level semantic, temporal, and spatial information~\citep{SS2A,viertola2026video,karchkhadze2026stereofoley,yoshida2026scene2sound}, but each prior effort is typically tied to a single audio format or covers only a subset of these dimensions. V2SA-Bench unifies these aspects through reference-free source-level evaluation of semantic, temporal, spatial, and acoustic alignment across stereo, binaural, and FOA audio. Leveraging recent advances in video segmentation~\citep{carion2026sam} and source separation~\citep{shi2025sam}, we establish cross-modal source correspondences for interpretable evaluation. While these tools serve as components, the contribution of V2SA-Bench lies in integrating them into a unified, reference-free, and source-aware evaluation protocol and in empirically validating its reliability and human alignment.
\section{V2SA-Bench}

\begin{figure}[t]
    \centering
    \includegraphics[width=0.85\linewidth]{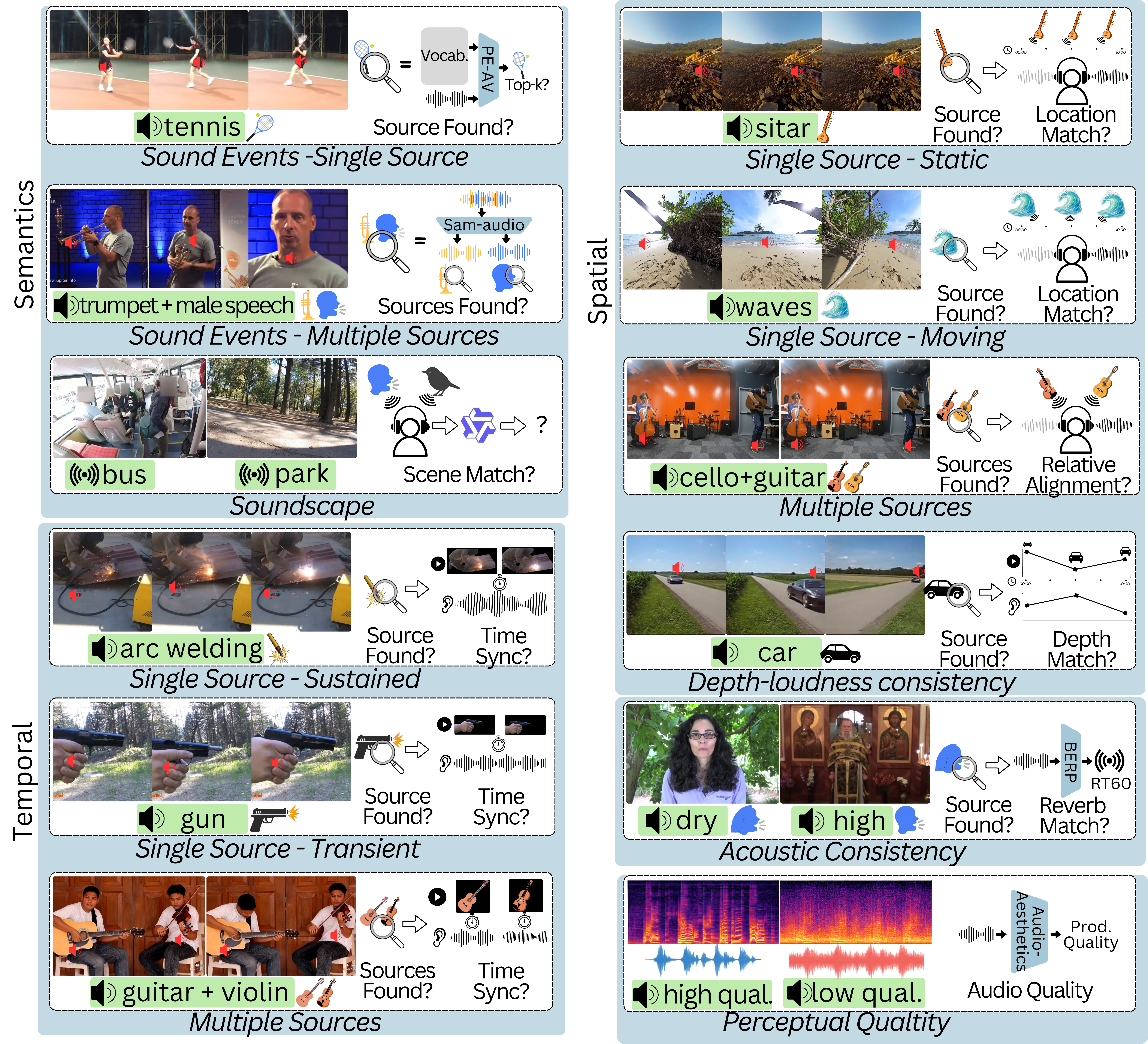}
    \caption{Examples and evaluation procedures for each dimension: semantic, temporal, spatial, acoustic, and perceptual quality. Each source-aware metric first verifies source presence, and only then assesses its timing, location, depth, or reverb match. In the Perceptual Quality panel, ``high qual.'' and ``low qual.'' denote high- and low-perceptual quality audio examples, respectively, scored by an audio-only production-quality metric (AudioBox Aesthetics~\citep{tjandra2025meta}).
    }
    \label{fig:metrics}
\end{figure}

We introduce V2SA-Bench to understand video-to-spatial audio generation models with a focus on source-aware evaluation. Given a perspective video, we consider two standard spatial audio formats, binaural and stereo. Binaural audio is recorded to simulate human experience, while stereo is recorded using left and right channels independently~\citep{wilkins2023two}.
%%%%
\subsection{V2SA-Bench taxonomy and definitions}
Our goal is to evaluate how realistically the generated audio aligns with the input video and its visible sound sources. To this end, we propose explicit audio-visual alignment metrics that quantify semantic, temporal, spatial, and acoustic consistency. Most of these metrics are source-aware, operating at the level of individual sound sources, while the semantic dimension additionally includes a scene-level metric that assesses the overall sound environment. For a more holistic evaluation, we additionally assess the perceptual quality of the generated audio using an audio-only metric.

We briefly define each evaluation dimension and its corresponding evaluation procedure below, with representative examples shown in Fig.~\ref{fig:metrics}.
Most of these procedures share a common count-based structure: for each clip, we first verify source presence and then check whether the corresponding alignment (e.g., temporal or spatial) satisfies a threshold, and we report the proportion of clips that pass as the metric.
The soundscape metric follows the same count-based scheme at the scene level, whereas the perceptual-quality metric instead reports an averaged score.
The cues used to compute our metrics (e.g., the target sources and their location) are derived from the video and its manual annotations (Sec.~\ref{ssec:data_curation}) rather than from reference audio.
\textit{Please refer to the supplementary for details of definition, full implementation details, prompt templates and examples.}

\noindent \textbf{Semantic alignment}. We evaluate semantic correspondence at two levels: \textit{sound events}, measuring whether visually expected sounds are present, and \textit{soundscape}, measuring consistency of the overall sound environment.
For sound events, we construct a global audio-visual vocabulary adapted from VGGSound~\citep{chen2020vggsound}, which also serves as a source-presence gate for other source-aware metrics.

\noindent $\bullet$ \textbf{\textit{Sound events}} -- \textbf{\textit{Single source}}. Given a target source (e.g., dog, car, or waves), we rank the vocabulary by PE-AV~\citep{vyas2026pushing} similarity with the generated audio and check whether the target appears in the top-$k$ ($k=5$), enabling training-free, zero-shot evaluation.

\noindent $\bullet$ \textbf{\textit{Sound events}} -- \textbf{\textit{Multiple sources}}. For multiple sources (e.g., trumpet and male speech), overlapping or otherwise, we isolate each source using SAM-Audio~\citep{shi2025sam} with class prompts and apply the same single-source semantic evaluation.

\noindent $\bullet$ \textbf{\textit{Soundscape}}. We adopt the urban acoustic-scene classes from~\citet{wang2021curated} (e.g., park, bus, or metro). For evaluation, we use \textit{Qwen3-Omni-30B-A3B-Thinking}~\citep{xu2025qwen3} model for closed-set classification and check whether the ground-truth scene appears among its top-$k$ ($k=3$) predictions.

%%%%%%%%%%%%%%%%%%%%%%%%%%%%%%%%%%%%%%%%%%%%%%%%%%%%%%%%%%%%%%%%%%
\noindent \textbf{Temporal alignment}. We evaluate audio-visual synchronization based on the number and temporal characteristics of sound sources. 
Single-source cases are further categorized into transient events with short, distinct energy peaks and sustained events with continuous activity. For source-wise evaluation, we isolate the target audio using SAM-Audio and mask and crop the video around the corresponding source, reducing interference from irrelevant audio-visual content.

\noindent $\bullet$ \textbf{\textit{Single source}} -- \textbf{\textit{Sustained}}. For sustained sources (e.g., arc welding or violin), we first verify source presence using our single-source semantic metric, then isolate the source with SAM-Audio and use Synchformer to measure synchronization with the 
masked visual source over time.

\noindent $\bullet$ \textbf{\textit{Single source}} -- \textbf{\textit{Transient}}. Unlike sustained sounds, transient events provide distinct temporal anchors for synchronization. Similar to the sustained case, for a transient event (e.g., gunshots or golf hits), we verify source presence and use Synchformer to measure synchronization.

\noindent $\bullet$ \textbf{\textit{Multiple sources}}.
We evaluate whether each visual source is temporally aligned with its corresponding generated sound. We first verify source presence using our multi-source semantic metric, then isolate each source with SAM-Audio and compute source-wise Synchformer scores using the corresponding masked and cropped video.

\noindent \textbf{Spatial alignment}. For an input video, we assess if the generated audio sources in the video are well aligned spatially in the horizontal plane.

\noindent $\bullet$ \textbf{\textit{Single source}} -- \textbf{\textit{Static}}.
Given a static sound source, this dimension evaluates if the sound source is present in the generated audio and subsequently spatially aligns with the visual position. For example, in Fig.~\ref{fig:metrics}, we check for if sitar is present and if it is on the right side. 
For binaural audio, we estimate azimuth by matching interaural level and phase difference (ILD/IPD) cues with head-related impulse response (HRIR)-derived templates using MagLS~\citep{schorkhuber2018binaural}; for stereo, we infer lateral direction from normalized left--right energy imbalance~\citep{chen2023nvas}.

\noindent $\bullet$ \textbf{\textit{Single source}} -- \textbf{\textit{Moving}}. 
For a moving sound source, we track the source relative to the camera using SAM3~\citep{carion2026sam} and evaluate whether the estimated audio localization follows its visual trajectory over time. For example, if a wave source moves from left to right in the video, the localized sound should exhibit a corresponding left-to-right trajectory.

\noindent $\bullet$ \textbf{\textit{Multiple sources}}. 
Given two static sound sources, we evaluate whether their relative visual positions are aligned in the generated audio. We apply SAM-Audio to each channel, average the separated outputs to verify source presence, and, if detected, use their left--right channel energies to estimate relative spatial alignment. For example, a cello appearing left of a guitar should also be localized to its left.

\noindent $\bullet$ \textbf{\textit{Depth--loudness}}\textbf{\textit{ consistency}}. 
For stable sound sources (e.g., cars, airplanes, or footsteps), audio should become louder as the source approaches the camera and quieter as it moves away. We first verify source presence, then measure whether relative root-mean-square (RMS) loudness is inversely correlated with visual source depth estimated using Depth Pro~\citep{bochkovskiy2025depth}.
%%%%%%%%%%%%%%%%%%%%%%%%%%%%%%%%%%%%%%%%%%%%

\noindent \textbf{Acoustic consistency}.
We evaluate whether the generated audio matches acoustic properties implied by the visual environment (e.g., speech in a church should exhibit high reverberation). We restrict this evaluation to speech, where reliable reverberation estimators are available. After verifying speech presence, we use BERP~\citep{wang2025berp} to estimate the reverberation time (i.e., RT60) and compare it with the visually expected dry or high reverberant condition.

\noindent \textbf{Perceptual quality}.
For a holistic evaluation of spatial audio generation, we complement audio-visual alignment metrics with the Production Quality (PQ) score from AudioBox Aesthetics~\citep{tjandra2025meta} to assess the perceptual quality of the generated audio.
\vspace{-3mm}
\subsection{Data curation}
\label{ssec:data_curation}
% %%%%%%%%%%%%%%%%%%%%%%%%%%%%%%%%
\begin{wrapfigure}{r}{0.45\textwidth}
    \vspace{-3mm}
    \centering
    \includegraphics[width=\linewidth]{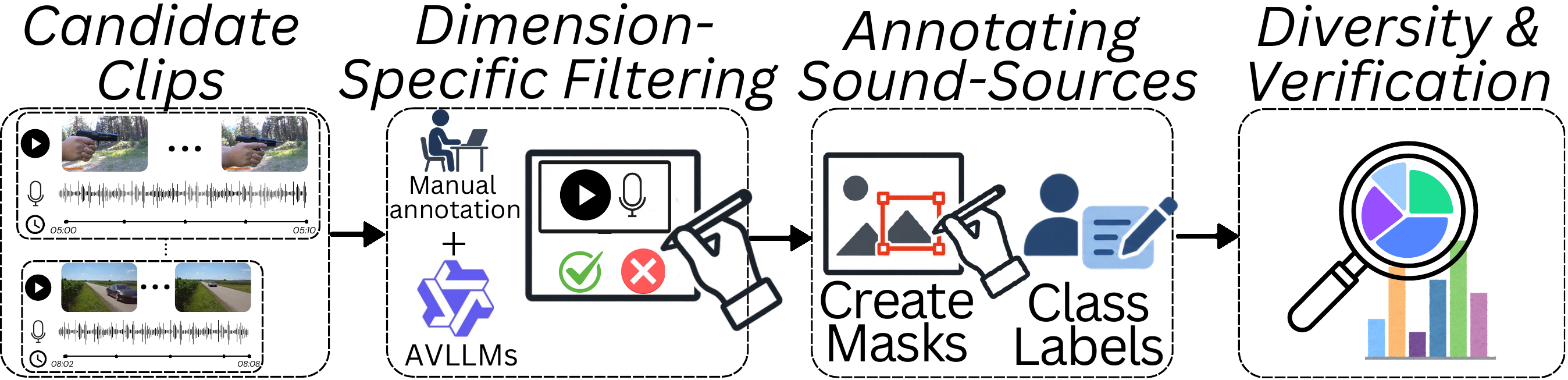}
    \caption{Data curation pipeline. Candidate clips are first extracted from source datasets, followed by dimension-specific filtering, source annotation, and diversity and quality verification.}
    \label{fig:data_curate}
    \vspace{-4mm}
\end{wrapfigure}
% %%%%%%%%%%%%%%%%%%%%%%%%%%%%%%
We curate the benchmark through a multi-stage pipeline ensuring annotation quality, dimension relevance, and data diversity.
Each dimension requires a different filtering process, though roughly following the pipeline shown in Fig.~\ref{fig:data_curate}. We also ensure that none of the clips come from the train set of training data of baseline models.
%%%%%%%%%%%%%%%%%%%%%%%%%%%%%%%
\begin{wraptable}{r}{0.45\textwidth}
\vspace{-6mm}
\centering
\caption{Source datasets used for candidate clip selection.}
\label{tab:candidate_selection}
\small
\resizebox{0.45\textwidth}{!}{
\begin{tabular}{p{0.18\textwidth}|p{0.29\textwidth}}
\toprule
\textbf{Dimension} & \textbf{Source Datasets} \\
\midrule
Semantics -- Single & UnAV-100\textcolor{red}{$^1$}, AVSync48\textcolor{red}{$^2$} \\
\midrule
Semantics -- Multi & YouTube-Stereo\textcolor{red}{$^3$}, FAIR-Play\textcolor{red}{$^4$}, MUSIC-Duet\textcolor{red}{$^5$}, AVSBench\textcolor{red}{$^6$}, UnAV-100\textcolor{red}{$^1$} \\
\midrule
Semantics -- Scene & TAU AV Scenes\textcolor{red}{$^7$} \\
\midrule
Temporal -- Single & UnAV-100\textcolor{red}{$^1$}, AVSync48\textcolor{red}{$^2$} \\
\midrule
Temporal -- Multi & YouTube-Stereo\textcolor{red}{$^3$}, FAIR-Play\textcolor{red}{$^4$}, MUSIC-Duet\textcolor{red}{$^5$}, AVSBench\textcolor{red}{$^6$}, UnAV-100\textcolor{red}{$^1$} \\
\midrule
Spatial -- Single & Sphere360\textcolor{red}{$^8$}, YT-Ambigen\textcolor{red}{$^9$} \\
\midrule
Spatial -- Multi & YouTube-Stereo\textcolor{red}{$^3$}, FAIR-Play\textcolor{red}{$^4$}, MUSIC-Duet\textcolor{red}{$^5$} \\
\midrule
Spatial -- Depth & UrbanSAS\textcolor{red}{$^{10}$}, VGGSound\textcolor{red}{$^{11}$} \\
\midrule
Acoustics (Reverb) & AVSpeech\textcolor{red}{$^{12}$} \\
\midrule
Perceptual Quality & UnAV-100\textcolor{red}{$^1$}, YouTube-Stereo\textcolor{red}{$^3$} \\
\bottomrule
\multicolumn{2}{l}{\tiny \textcolor{red}{$^1$}~\citet{geng2023dense}, \textcolor{red}{$^2$}~\citet{zhang2025scaling}, \textcolor{red}{$^3$}~\citet{sato2025audio}, \textcolor{red}{$^4$}~\citet{gao20192},} \\
\multicolumn{2}{l}{\tiny \textcolor{red}{$^5$}~\citet{zhao2018sound}, \textcolor{red}{$^6$}~\citet{zhou2022audio}, 
\textcolor{red}{$^7$}~\citet{wang2021curated}, \textcolor{red}{$^8$}~\citet{liu2025omniaudio},
} \\
\multicolumn{2}{l}{\tiny \textcolor{red}{$^9$}~\citet{kim2025visage},
\textcolor{red}{$^{10}$}~\citet{fuentes2022urban}, 
\textcolor{red}{$^{11}$}~\citet{chen2020vggsound}, 
\textcolor{red}{$^{12}$}~\citet{ephrat2018looking}
} \\
\end{tabular}
}
\vspace{-4mm}
\end{wraptable}
%%%%%%%%%%%%%%%%%%%%%%%%%%%%%%%

\noindent \textbf{Candidate clip selection}. We curate candidate clips from the source datasets summarized in Tab.~\ref{tab:candidate_selection}.
For single-source spatial clips, we follow ViSAudio~\citep{zhang2025visaudio} and crop perspective views from 360° video, decoding their FOA audio to stereo/binaural.
This provides controlled source placement in diverse environments.
For multi-source spatial clips, since it is difficult to place multiple sources within the cropped field of view, we instead use native stereo/binaural recordings.

\noindent \textbf{Dimension-specific filtering}.
For sound-event dimensions, we require the sound source to be visually evident and unambiguous; for multi-source cases, the sources must additionally be visually non-overlapping. We filter candidates using \textit{Qwen3-Omni-30B-A3B-Thinking}~\citep{xu2025qwen3}, followed by manual inspection. For acoustic-scene and soundscape evaluation, we select clips with clearly identifiable environments (e.g., outdoor spaces or reverberant interiors such as churches), whereas distance clips contain sources that clearly approach or recede from a relatively stationary camera.

\noindent \textbf{Annotations}. We adapt a controlled source vocabulary from VGGSound by merging action-specific labels into common source classes (e.g., \emph{dog barking}, \emph{dog howling} $\rightarrow$ \emph{dog}), yielding 162 classes across humans, animals, vehicles, instruments, household objects, and natural sounds, and manually annotate the active sources in each clip. For source-level spatial and temporal evaluation, we obtain per-source visual masks using SAM3; for classes not well represented by masks (e.g., dart throwing or fireworks), we use bounding boxes instead.

\noindent \textbf{Verification and diversity filtering}. Finally, we manually verify semantic correctness, audio-visual consistency, and annotation quality, including source labels and masks, while filtering the set to maintain broad source and scene diversity. This process yields 767 unique perspective videos, with detailed benchmark statistics provided in the appendix.
\section{Experiments}

\subsection{Experimental setup}

\noindent \textbf{Baselines.}
We compare against publicly available video-to-spatial audio generation methods spanning native two-channel and richer multichannel outputs. For native stereo generation, we evaluate AudioX~\citep{tian2026audiox}, ThinkSound~\citep{liu2026thinksound}, PrismAudio~\citep{liu2026prismaudio}, and SpatialSonic~\citep{sun2025both}. For binaural generation, we additionally construct a cascaded baseline of two stages: MMAudio~\citep{cheng2025mmaudio} for video-to-mono audio generation, followed by Mono2Binaural (M2B)~\citep{gao20192} for video-guided audio spatialization.
We also include methods that generate richer spatial formats and decode them to two-channel audio: ViSAGe~\citep{kim2025visage}, an end-to-end model, generates FOA, whereas SEE-2-SOUND~\citep{dagli2025see}, a cascaded pipeline, generates 5.1-channel audio. Both are converted to stereo and binaural for evaluation.
Since our metrics are reference-free, we additionally evaluate ground-truth audio. % and a random baseline.
We restrict comparisons to models with publicly available checkpoints, excluding methods such as StereoFoley~\citep{karchkhadze2026stereofoley}, ViSAudio~\citep{zhang2025visaudio} and SwanSphere~\citep{lei2026towards}, whose checkpoints or training pipelines are not publicly released.
We infer each of these models only using the video. Any model with input text is kept empty(``''), so that, for fair comparison, each model only infers from video input.
% 
%%%%%%%%%%%%%%%%%%%%%%%%%%%%%%%%%%%%%%%%%%%%%%%%%%%%%%%%%%%%%%%%%%%%%
%%%%%%%%%%%%%%%%%%%%%%%%%%%%%%%%%%%%%%%%%%%%%%%%%%%%%%%%%%%%%%%%%%%%%
\begin{table*}[t]
\centering
\caption{Comparison of models across our evaluation metrics. Metrics are organized into 5 dimensions (semantics, temporal, spatial, acoustics, and perceptual quality), each further divided into fine-grained sub-categories. GT denotes the ground-truth reference, and the best result in each task (excluding GT) is shown in \textbf{bold}. Higher is better for all metrics.
}
\label{tab:main_results}
\renewcommand{\arraystretch}{1.25}
\setlength{\tabcolsep}{4pt}
\resizebox{\textwidth}{!}{%
%%%%%%%%%%%%%%%%%%%%%%%%%%%%%%%%%%%%%%%%%%%%%%
\begin{tabular}{l|ccc|ccc|cccc|c|c} \toprule \multirow[c]{3}{*}{\textbf{Method}}
& \multicolumn{3}{c|}{\textbf{Semantics}}
& \multicolumn{3}{c|}{\textbf{Temporal}}
& \multicolumn{4}{c|}{\textbf{Spatial}}
& \multirow[c]{3}{*}{\textbf{Acoustics}}
& \multirow[c]{3}{*}{\makecell{\textbf{Perceptual}\\\textbf{Quality}}}
\\
\cmidrule(lr){2-4}
\cmidrule(lr){5-7}
\cmidrule(lr){8-11}
& \makecell{Single\\(Events)}
& \makecell{Multi.\\(Events)}
& Soundscape
& Sustained
& Transient
& Multi
& \makecell{Static\\(Single)}
& \makecell{Moving\\(Single)}
& Multi.
& Depth
&
&
\\
\midrule
AudioX\color{red}{$^1$} & 87.9 & 50.9 & \textbf{56.0} & 32.3 & 33.9 & 15.4 & 16.1 & 12.1 & \textbf{16.2} & \textbf{40.8} & 37.0 & \textbf{6.5} \\
ThinkSound\color{red}{$^2$} & 83.3 & 57.0 & 37.0 & 32.3 & 35.7 & 33.3 & 25.0 & 9.4 & 15.1 & 28.6 & 44.9 & 6.1 \\
PrismAudio\color{red}{$^3$} & 26.8 & 26.6 & 35.0 & 0.0 & 5.3 & 11.2 & 4.4 & 0.4 & 9.3 & 27.6 & 30.3 & 5.5 \\
SpatialSonic\color{red}{$^4$} & 70.3 & 10.5 & 49.0 & 11.2 & 23.2 & 8.1 & 19.1 & 4.7 & 4.6 & 39.4 & 25.8 & 5.4 \\
SEE-2-SOUND (Stereo)\color{red}{$^5$} & 16.6 & 16.4 & 30.0 & 2.8 & 7.1 & 6.8 & 4.4 & 7.0 & 3.4 & 1.2 & 6.7 & 5.2 \\
SEE-2-SOUND (Binaural)\color{red}{$^5$} & 16.6 & 16.6 & 30.0 & 1.4 & 0.0 & 7.1 & 1.4 & 6.3 & 0.0 & 0.7 & 7.8 & 4.6 \\
ViSAGe (Stereo)\color{red}{$^6$} & 29.6 & 27.4 & 37.0 & 2.8 & 5.3 & 7.9 & 17.6 & 8.8 & 3.4 & 15.1 & 22.4 & 5.4 \\
ViSAGe (Binaural)\color{red}{$^6$} & 32.4 & 26.6 & 37.0 & 4.2 & 0.0 & 8.5 & 16.1 & 13.2 & 10.4 & 16.5 & 17.9 & 5.2 \\
{MMAudio\color{red}{$^7$}} + M2B\color{red}{$^8$} & \textbf{98.1} & \textbf{70.0} & 50.0 & \textbf{66.2} & \textbf{75.0} & \textbf{42.0} & \textbf{26.4} & \textbf{25.3} & 15.1 & 33.8 & \textbf{48.3} & 6.1 \\
% {Hunyuan-V2A\color{red}{$^9$}} + M2B\color{red}{$^8$} & - & - & - & - & - & - & - & - & - & - & - & - \\
\midrule
\demph{GT} & \demph{99.0} & \demph{77.7} & \demph{78.0} & \demph{70.4} & \demph{64.2} & \demph{65.2} & \demph{77.2} & \demph{70.2} & \demph{68.6} & \demph{72.8} & \demph{95.5} & \demph{6.3} \\
% \demph{Random} & \demph{4.8} & \demph{4.9} & \demph{30.0} & \demph{1.3} & \demph{0.9} & \demph{1.4} & \demph{1.8} & - & - & - & - & - \\
\bottomrule
\multicolumn{13}{l}{\scriptsize \textcolor{red}{$^1$}~\citet{tian2026audiox}, \textcolor{red}{$^2$}~\citet{liu2026thinksound}, \textcolor{red}{$^3$}~\citet{liu2026prismaudio}, \textcolor{red}{$^4$}~\citet{sun2025both}, \textcolor{red}{$^5$}~\citet{dagli2025see}, \textcolor{red}{$^6$}~\citet{kim2025visage}, \textcolor{red}{$^7$}~\citet{cheng2025mmaudio}, \textcolor{red}{$^8$}~\citet{gao20192}}
\end{tabular}
}
\end{table*}
%%%%%%%%%%%%%%%%%%%%%%%%%%%%%%%%%%%%%%%%%%%%%%%%%%%%%%%%%%%%%%%%%%%%%%
%%%%%%%%%%%%%%%%%%%%%%%%%%%%%%%%%%%%%%%%%%%%%%%%%%%%%%%%%%%%%%%%%%%%%
\subsection{Experimental results}

We show our experimental results in Tab.~\ref{tab:main_results}. Our key observations are summarized below.

\noindent\textbf{Video-to-mono generation followed by video-guided spatialization shows a strong trend}. The cascaded MMAudio + M2B method achieves the best results on 8 of 12 dimensions, ahead of native stereo and richer spatial-format generation models. Notably, being cascaded alone is not sufficient: SEE-2-SOUND, which grounds visual sources in 3D, generates per-source mono audio with CoDi~\citep{tang2023any}, and spatializes it via room simulation, is also cascaded yet remains among the lowest-scoring. This suggests that the advantage stems from a high-quality mono first stage whose fidelity the video-guided spatialization preserves while aligning the sound with the video.

\noindent\textbf{Spatial reasoning remains a major bottleneck}. The gap between GT and the best baseline is highly uneven across dimensions: it is negligible for single-source semantics (99.0 vs. 98.1) but drastic for spatial alignment (e.g., 77.2 vs. 26.4 for static and 68.6 vs. 16.2 for multi-source). This reveals spatial alignment as the most challenging dimension, showing that current models struggle to localize sound sources according to the visual scene. Moreover, richer spatial-audio formats bring no clear advantage: FOA- and 5.1-based models (ViSAGe, SEE-2-SOUND) perform on par with, and sometimes below, native stereo/binaural models on spatial dimensions.

\noindent\textbf{On the ground-truth reference scores}.  Ground-truth audio does not reach 100\% on several dimensions, such as multi-source semantics and transient synchronization, due to strict source-level criteria, real-world ambiguities (e.g., overlapping sources and subtle transients), and evaluator limitations.
To assess robustness, we vary semantic top-$k$ and synchronization thresholds (refer Fig.~\ref{fig:gt_ceil}). While more permissive thresholds improve absolute GT scores, baseline rankings remain largely stable, indicating that our comparative conclusions are robust to threshold choices. Thus, GT serves as a practical reference rather than an absolute ceiling for our reference-free evaluation protocol.

\noindent \textbf{Increasing the number of sources hurts performance}. Moving from single to multiple sources degrades alignment across all dimensions. Taking the overall best model MMAudio + M2B as a representative example, its scores drop from single- to multi-source settings for semantics (98.1 $\rightarrow$ 70.0), temporal (66.2/75.0 for sustained/transient $\rightarrow$ 42.0), and spatial (26.4/25.3 for static/moving $\rightarrow$ 15.1) alignment. The same trend holds for the other models as well. This indicates that strong single-source performance does not extend to multi-source scenes, highlighting the need for source-wise evaluation rather than extrapolating scene-level capability from a collection of individual sources.

\noindent \textbf{Higher perceptual quality does not imply better alignment}. AudioX achieves the best PQ (6.5) but only mid-tier alignment scores (e.g., 16.1 on static spatial, 32.3 on sustained temporal), whereas MMAudio + M2B, with a lower PQ (6.1), leads in most alignment dimensions (26.4 and 66.2, respectively). Perceptual quality and audio-visual alignment are thus distinct capabilities.

\begin{figure}[t]
    \centering
    \includegraphics[width=\linewidth]{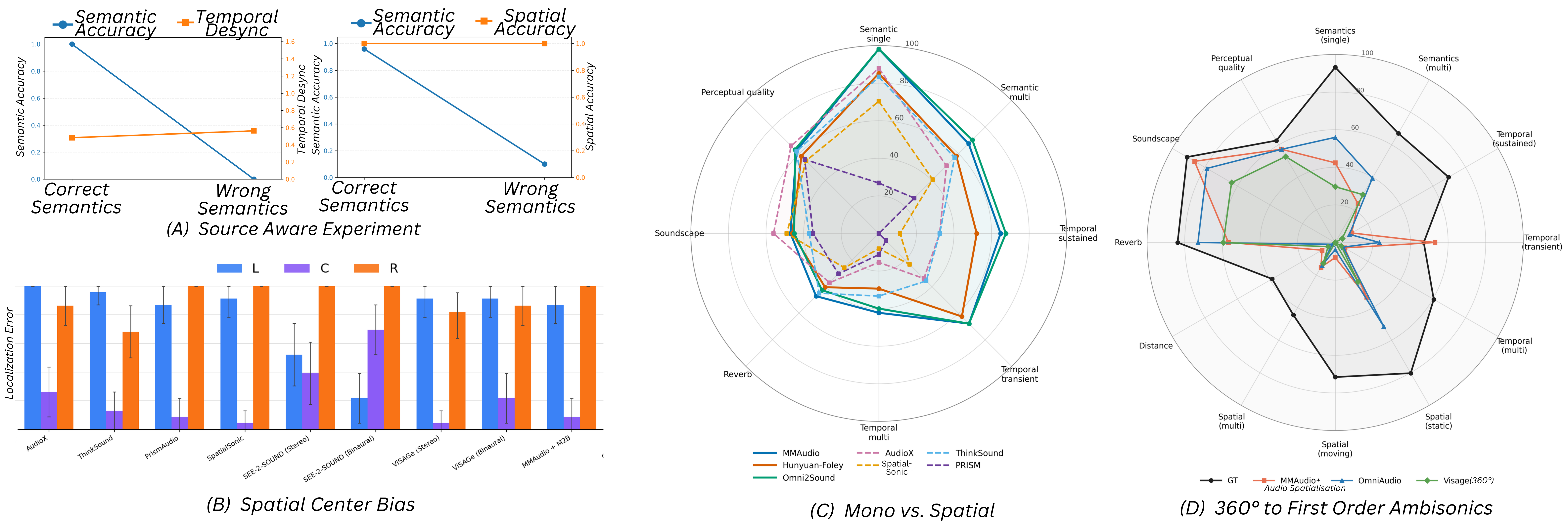}
    \caption{Additional analyses.
    (A) Under semantic perturbations, temporal and spatial scores stay high while semantic accuracy drops, showing that source-presence checking is essential. (B) Localization error is lowest at the center and grows toward the sides, revealing a spatial center bias. (C) Mono vs. spatial generation across non-spatial metrics. (D) Results on the 360° $\rightarrow$ FOA setting.
    }
    \label{fig:add_exp}
\end{figure}
%%%%%%%%%%%%%%%%%%%%%%%%%%%%%%%%
\begin{table*}[t]
\centering
\caption{Reliability experiments. We compare our evaluator models and methods used in each dimension. Semantics dimension (a)-(c) compares accuracy($\uparrow$); Temporal dimension (d)-(f) compares audio-visual synchronisation ($\downarrow$), where lower is better; Spatial dimension (g) compares spatial accuracy($\uparrow$); Depth-loudness dimension (h) compares alignment score($\uparrow$); Acoustics (i) compares accuracy ($\uparrow$); and Perceptual Quality (j) compares pairwise accuracy ($\uparrow$) (gives human preference).}
\label{tab:evaluator_comparison}
\vspace{3pt}

\resizebox{\textwidth}{!}{%
\begin{minipage}{\textwidth}
\centering
\scriptsize
\setlength{\tabcolsep}{2.2pt}
\renewcommand{\arraystretch}{1.02}

\newcommand{\evalpanel}[3]{%
  \begin{minipage}[t]{#1}
  \vspace{0pt}\centering
  \textbf{#2}\par\vspace{2pt}
  #3
  \end{minipage}%
}

\evalpanel{0.24\textwidth}{(a) Semantics-Single}{%
  \begin{tabular}{@{}ll@{}}
  \toprule
  Model & Acc.(\%) \\
  \midrule
  PE-AV$^1$(ours) & \textbf{99.1} \\
  CLAP$^2$ & 86.1 \\
  IB$^3$  & 93.5 \\
  \bottomrule
  \end{tabular}
}%
\hfill
\evalpanel{0.31\textwidth}{(b) Semantics-multi}{%
  \setlength{\tabcolsep}{1.4pt}
  \begin{tabular}{@{}lccc@{}}
  \toprule
  Models & \shortstack{SAM-Audio$^4$\\(ours)} & Solo-Audio$^5$ & CLAP-Sep$^6$ \\
  \midrule
  PE-AV$^1$(ours) & \textbf{77.7} & 73.6 & 69.0 \\
  CLAP$^2$ & 67.8 & 48.4 & 62.8 \\
  IB$^3$   & 58.7 & 55.0 & 55.4 \\
  \bottomrule
  \end{tabular}
}%
\hfill
\evalpanel{0.31\textwidth}{(c) Soundscape}{%
  \begin{tabular}{@{}lr@{}}
  \toprule
  Model & Acc.(\%) \\
  \midrule
  Qwen-3-Omni$^7$(ours) & \textbf{78.0} \\
  Gemma-4-12B$^8$ & 30.0 \\
  Audio Flamingo Next$^9$ & 56.0 \\
  \bottomrule
  \end{tabular}
}%

\par\vspace{4pt}

\evalpanel{0.31\textwidth}{(d) Temporal-Sustained}{%
  \begin{tabular}{@{}r|rr@{}}
  \toprule
  Offset & \shortstack{Global-scene$^{10}$\\Sync} & \shortstack{Source-wise\\(ours)} \\
  \midrule
  $-2$ & 1.197 & 0.256 \\
  $-1$ & 0.527 & 0.232 \\
  $0$  & 0.300 & 0.528 \\
  $+1$ & 0.477 & 0.280 \\
  $+2$ & 1.108 & 0.285 \\
  \midrule
  Mean & 0.721 & \textbf{0.316} \\
  \bottomrule
  \end{tabular}
}%
\hfill
\evalpanel{0.31\textwidth}{(e) Temporal-Transient}{%
  \begin{tabular}{@{}r|rr@{}}
  \toprule
  Offset & \shortstack{Global-scene$^{10}$\\Sync} & \shortstack{Source-wise\\(ours)} \\
  \midrule
  $-2$ & 1.188 & 0.232 \\
  $-1$ & 0.534 & 0.234 \\
  $0$  & 0.270 & 0.396 \\
  $+1$ & 0.634 & 0.286 \\
  $+2$ & 1.386 & 0.364 \\
  \midrule
  Mean & 0.802 & \textbf{0.302} \\
  \bottomrule
  \end{tabular}
}%
\hfill
\evalpanel{0.31\textwidth}{(f) Temporal-multi}{%
  \begin{tabular}{@{}r|rr@{}}
  \toprule
  Offset & \shortstack{Global-scene$^{10}$\\Sync} & \shortstack{Source-wise\\(ours)} \\
  \midrule
  $-2$ & 1.408 & 0.264 \\
  $-1$ & 0.642 & 0.224 \\
  $0$  & 0.271 & 0.487 \\
  $+1$ & 0.631 & 0.296 \\
  $+2$ & 1.476 & 0.369 \\
  \midrule
  Mean & 0.885 & \textbf{0.328} \\
  \bottomrule
  \end{tabular}
}

\par\vspace{4pt}

\evalpanel{0.29\textwidth}{(g) Spatial accuracy}{%
  \begin{tabular}{@{}lcc@{}}
  \toprule
  Setting & Ours (\%) & BAT$^{11}$ (\%) \\
  \midrule
  Single static & \textbf{97.8} & 56.7 \\
  Single moving & \textbf{82.9} & 55.0 \\
  Multi-source  & \textbf{75.6} & 49.4 \\
  \bottomrule
  \end{tabular}
}%
\hfill
\evalpanel{0.20\textwidth}{(h) Depth-loudness}{%
  \begin{tabular}{@{}lr@{}}
  \toprule
  Method & Score \\
  \midrule
  Ours & \textbf{73.9} \\
  BAT$^{11}$  & 11.3 \\
  \bottomrule
  \end{tabular}
}%
\hfill
\evalpanel{0.23\textwidth}{(i) Acoustics}{%
  \begin{tabular}{@{}lr@{}}
  \toprule
  Evaluator & Acc. \\
  \midrule
  BERP$^{12}$(ours)      & \textbf{100.0} \\
  Rec-RIR$^{13}$     & 83.1 \\
  blind\_rt60$^{14}$ & 59.6 \\
  \bottomrule
  \end{tabular}
}%
\hfill
\evalpanel{0.25\textwidth}{(j) Perceptual quality}{%
  \begin{tabular}{@{}lr@{}}
  \toprule
  Evaluator & Acc. \\
  \midrule
  AudioBox$^{15}$(ours)  & \textbf{100.0} \\
  PAM$^{16}$       & 75.0 \\
  NISQA$^{17}$     & 50.0 \\
  DNSMosPRo$^{18}$ & 35.0 \\
  \bottomrule
  \end{tabular}
}

\par\vspace{2pt}

\begin{minipage}{\textwidth}
  \tiny
  \raggedright
  \textcolor{red}{\textsuperscript{1}}~\citet{vyas2026pushing},
  \textcolor{red}{\textsuperscript{2}}~\citet{liu2026thinksound},
  \textcolor{red}{\textsuperscript{3}}~\citet{girdhar2023imagebind},
  \textcolor{red}{\textsuperscript{4}}~\citet{shi2025sam},
  \textcolor{red}{\textsuperscript{5}}~\citet{helin2024soloaudio},
  \textcolor{red}{\textsuperscript{6}}~\citet{ma2024clapsep},
  \textcolor{red}{\textsuperscript{7}}~\citet{xu2025qwen3},
  \textcolor{red}{\textsuperscript{8}}~\citet{team2026gemma},
  \textcolor{red}{\textsuperscript{9}}~\citet{ghosh2026audio},
  \textcolor{red}{\textsuperscript{10}}~\citet{iashin2024synchformer},
  \textcolor{red}{\textsuperscript{11}}~\citet{zheng2024bat},
  \textcolor{red}{\textsuperscript{12}}~\citet{wang2025berp},
  \textcolor{red}{\textsuperscript{13}}~\citet{wang2025recrirmonauralblindroom},
  \textcolor{red}{\textsuperscript{14}}~\citet{ratnam2003blind},
  \textcolor{red}{\textsuperscript{15}}~\citet{tjandra2025meta},
  \textcolor{red}{\textsuperscript{16}}~\citet{deshmukh2024pam},
  \textcolor{red}{\textsuperscript{17}}~\citet{mittag2021nisqa},
  \textcolor{red}{\textsuperscript{18}}~\citet{cumlin24_interspeech},
\end{minipage}

\end{minipage}%
}
\end{table*}
%%%%%%%%%%%%%%%%%%%%%%%%%%%%%%%%
\subsection{Additional analysis}\label{sec:additional_analysis}
\noindent \textbf{Source presence checking is important}. Existing audio-visual temporal and spatial alignment metrics are primarily designed for recorded audio, where the presence of the correct sound source is typically assumed. Directly applying these metrics to generated audio can therefore reward outputs that are temporally or spatially aligned but semantically incorrect.
To illustrate this issue, we construct two controlled perturbation settings. For temporal alignment, we use ControlFoley~\citep{yang2026controlfoleyunifiedcontrollablevideotoaudio} to generate audio with incorrect source semantics while preserving similar temporal dynamics (e.g., generating bicycle-acceleration audio for a video of a crying baby). For spatial alignment, we replace a spatially localized sound source with a semantically different source while preserving its spatial position. As shown in Fig.~\ref{fig:add_exp}~(A), semantic accuracy drops substantially under these perturbations, while temporal and spatial alignment scores remain nearly unchanged. This demonstrates that high temporal or spatial alignment alone does not guarantee that the correct sound source is present, motivating source-presence checking before source-level evaluation.

\noindent \textbf{Evaluator reliability}. Since V2SA-Bench relies on automatic evaluator components, we further validate the selected evaluators against alternative choices in Table 4. For each evaluation dimension, we compare the evaluator used in V2SA-Bench with representative alternatives on ground-truth or controlled validation data. The selected evaluators perform best among the tested alternatives in these validation settings, supporting their use in our benchmark. At the same time, V2SA-Bench should be interpreted as a reference-free diagnostic protocol whose accuracy is bounded by the underlying recognition, separation, synchronization, localization, and acoustic-estimation models.

\noindent \textbf{Spatial center bias}.
Figure~\ref{fig:add_exp}~(B) reveals a pronounced center bias across current models: localization error is consistently lowest for centrally located sources and grows toward the left and right. This suggests that models often default toward central spatial predictions rather than faithfully following the visual source direction. A likely contributor is directional imbalance in the training data, highlighting source-direction rebalancing as a concrete future step for improving spatial generation.

\noindent \textbf{Spatial vs. mono audio generation}.
Our source-aware metrics can be used for mono audio generation except the spatial metrics. On these metrics, we compare native stereo generation and mono audio generation methods in Fig.~\ref{fig:add_exp}~(C). We observe that mono audio generation models perform better in most dimensions except perceptual quality and soundscape. This supports our earlier observation that the strength of cascaded pipelines stems from a high-quality mono first stage, whose alignment is preserved by the subsequent video-guided spatialization.

\begin{figure}[t]
    \centering
    \includegraphics[width=\linewidth]{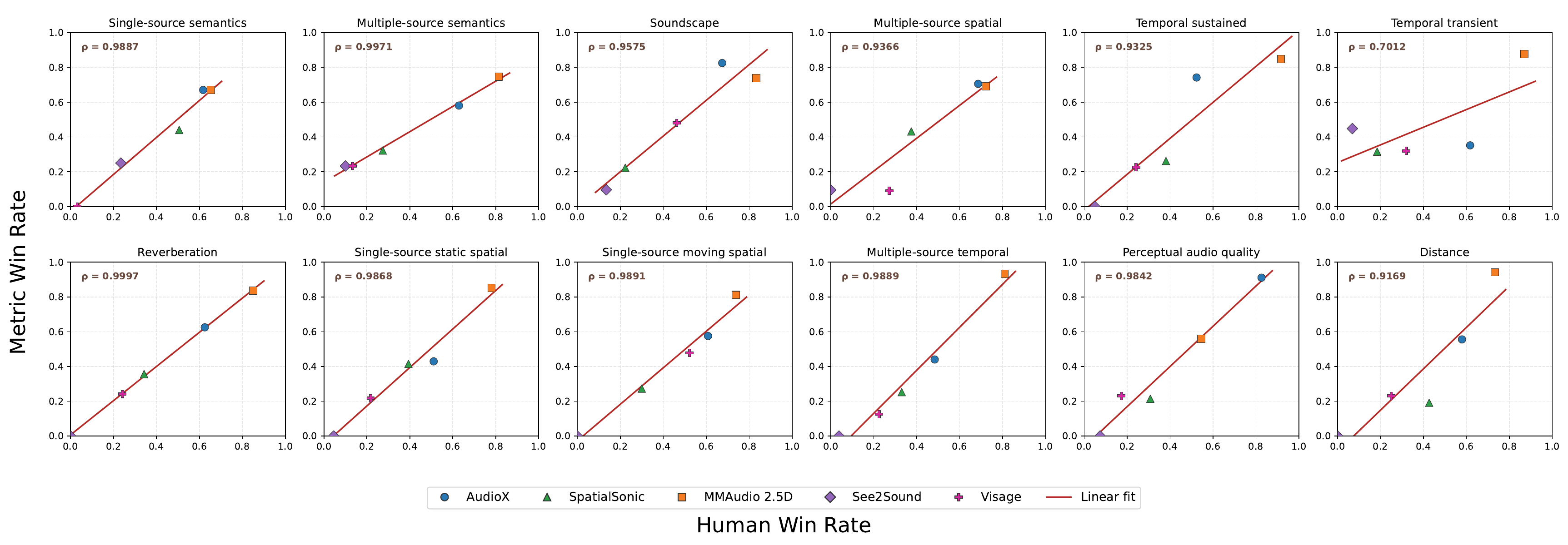}
    \vspace{-6mm}
    \caption{Human study. Per-dimension win rates from our automatic metrics (\textit{Metric Win Rate}) and from blinded human pairwise comparisons (\textit{Human Win Rate}). The strong agreement (most dimensions $>$ 0.9 correlation) indicates that our metrics align well with human preference.
    }
    \label{fig:human_study}
\end{figure}

\noindent \textbf{Human evaluation}.
We conducted blinded pairwise comparisons across all 12 evaluation dimensions and five models: AudioX, SpatialSonic, MMAudio + M2B,  ViSAGe(stereo) and See-2-Sound(stereo). In each trial, participants listened to the outputs from two anonymized models and selected which better satisfied the specified evaluation criterion, with additional options for \emph{Both are similar} and \emph{Can't judge}. Model and trial order were randomized, as was dimension order in the combined study. Participants were instructed to use headphones and evaluate only the requested attribute.
After data collection, we retained participants who completed all 22 comparison trials for per dimension: 15 among AudioX, SpatialSonic, and MMAudio 2.5D, and seven additional comparisons involving SEE-2-SOUND and ViSAGe. Overall, 20 unique participants contributed 168 completed participant–dimension sessions across 12 dimensions. Human preferences were scored as win $=1$, loss $=0$, and tie $=0.5$, excluding \emph{Can't judge} responses. Human and metric win rates were computed over the same evaluable comparisons. 
As shown in Fig.~\ref{fig:human_study}, in our human study, human preferences showed strong agreement with our automatic metrics, with most dimensions achieving correlations above $0.9$.
\subsection{Extending to 360° videos to FOA}

Our source-aware metrics extend naturally beyond perspective videos to the 360° video $\rightarrow$ FOA task. The metrics remain unchanged, except that spatial localization is performed with a source localization model~\citep{shimada2025open} on the FOA signal, in place of the stereo/binaural estimators used for perspective videos.
We source 360° clips from Sphere360~\citep{liu2025omniaudio}, YT-Ambigen~\citep{kim2025visage}, YT360-EyeTracking~\citep{cokelek2025spherical}, and YT-All~\citep{morgado2018self}, and apply a curation process similar to the perspective setting, yielding 507 unique 360° videos.

We evaluate two native FOA audio generation models, OmniAudio~\citep{liu2025omniaudio} and ViSAGe~\citep{kim2025visage}. OmniAudio natively operates on 360° video, whereas the original ViSAGe takes a perspective video with a target direction as input. For a fair comparison under our 360° setting, we follow the OmniAudio setup and train ViSAGe directly on 360° video.
We additionally consider cascaded baselines that decompose spatial audio generation into video-to-mono generation followed by audio spatialization. For the first stage, the equirectangular 360° video is resized to the input resolution required by the video-to-mono audio generation model~\citep{cheng2025mmaudio}. For spatialization, we employ SalViT~\citep{cokelek2025spherical} to estimate the salient source location at 100-ms intervals and spatialize the generated mono audio according to these time-varying locations.

% We show the experimental results in Fig.~\ref{fig:add_exp}~(D). Overall, the 360° $\rightarrow$ FOA setting exhibits trends consistent with the perspective setting. The cascaded video-to-mono generation followed by audio spatialization again shows a strong trend, while the end-to-end OmniAudio achieves competitive performance. Spatial reasoning remains the primary bottleneck, with a particularly pronounced degradation on moving sources. The remaining observations, i.e., performance drops from single to multiple sources and higher perceptual quality not implying better alignment, also hold in this setting.
We show the results in Fig.~\ref{fig:add_exp}~(D). The 360° $\rightarrow$ FOA setting largely mirrors the perspective setting: the cascaded mono-then-spatialize pipeline again performs strongly, while the end-to-end OmniAudio remains competitive. Spatial reasoning stays the main bottleneck, most notably on moving sources, and the other observations also hold here.

\section{Conclusion}

We introduced V2SA-Bench, a source-aware benchmark for video-to-spatial audio generation, comprising 1,274 human-verified clips across two video and three spatial-audio formats. We propose 12 reference-free, human-aligned metrics spanning five dimensions, enabling fine-grained source-level evaluation. Across 13 baselines, we provide the first source-wise quantitative characterization of spatial center bias and identify spatial alignment as a key bottleneck. We further find that cascaded models generally outperform end-to-end models due to stronger mono generation, while perceptual quality does not necessarily imply better alignment. Performance declines with increasing source complexity, highlighting the importance of source-aware evaluation and source-presence verification. We hope V2SA-Bench guides future advances in spatial audio generation.

\noindent \textbf{Limitations and future work}. 
Although our evaluators outperform the alternatives (Table~\ref{tab:benchmark_comparison}), their accuracy is bounded by the underlying models, which may improve with future advances.
Our scope is also limited to video-conditioned generation; extending to text-conditioned inputs and broadening domain coverage, particularly speech and music, is a promising future work.

\bibliography{iclr2027_conference}

@STRING{PROC              = "Proc. "}

@STRING{arXiv             = "{arXiv preprint arXiv}"}

@STRING{AAAI              = "AAAI Conference on Artificial Intelligence (AAAI)"}

@STRING{CVPR              = "IEEE Conference on Computer Vision and Pattern Recognition (CVPR)"}

@STRING{IEEE_ICASSP       = "{IEEE} Int. Conf. Acoust., Speech, Signal Process. ({ICASSP})"}

@INPROCEEDINGS{gui2024fadtk,
  author={Gui, Azalea and Gamper, Hannes and Braun, Sebastian and Emmanouilidou, Dimitra},
  booktitle=Proc#IEEE_ICASSP, 
  title={Adapting Frechet Audio Distance for Generative Music Evaluation}, 
  year={2024},
  volume={},
  number={},
  pages={1331-1335},}

@article{zhao2026foleyspace,
  title={Foleyspace: Vision-aligned binaural spatial audio generation},
  author={Zhao, Lei and Chen, Rujin and Zhang, Chi and Zhang, Xiao-Lei and Li, Xuelong},
  journal={IEEE Transactions on Multimedia},
  year={2026},
  publisher={IEEE}
}

@article{lei2026towards,
  title={Towards Streaming Synchronized Spatial Audio Generation via Autoregressive Diffusion Transformer},
  author={Lei, Ke and Zhang, Yu and Pan, Changhao and Pu, Xueyi and Guo, Wenxiang and Li, Ruiqi and Zhao, Zhou},
  journal={arXiv preprint arXiv:2605.30940},
  year={2026}
}

@article{tang2023any,
  title={Any-to-any generation via composable diffusion},
  author={Tang, Zineng and Yang, Ziyi and Zhu, Chenguang and Zeng, Michael and Bansal, Mohit},
  journal={Advances in Neural Information Processing Systems},
  volume={36},
  pages={16083--16099},
  year={2023}
}

@inproceedings{fuentes2022urban,
  title={Urban sound \& sight: Dataset and benchmark for audio-visual urban scene understanding},
  author={Fuentes, Magdalena and Steers, Bea and Zinemanas, Pablo and Rocamora, Martin and Bondi, Luca and Wilkins, Julia and Shi, Qianyi and Hou, Yao and Das, Samarjit and Serra, Xavier and others},
  booktitle={ICASSP 2022-2022 IEEE International Conference on Acoustics, Speech and Signal Processing (ICASSP)},
  pages={141--145},
  year={2022},
  organization={IEEE}
}

@article{kushwaha2023multimodal,
  title={A multimodal prototypical approach for unsupervised sound classification},
  author={Kushwaha, Saksham Singh and Fuentes, Magdalena},
  journal={arXiv preprint arXiv:2306.12300},
  year={2023}
}

@inproceedings{kushwaha2025diff,
  title={Diff-sage: End-to-end spatial audio generation using diffusion models},
  author={Kushwaha, Saksham Singh and Ma, Jianbo and Thomas, Mark RP and Tian, Yapeng and Bruni, Avery},
  booktitle={ICASSP 2025-2025 IEEE International Conference on Acoustics, Speech and Signal Processing (ICASSP)},
  pages={1--5},
  year={2025},
  organization={IEEE}
}

@inproceedings{zhao2018sound,
  title={The sound of pixels},
  author={Zhao, Hang and Gan, Chuang and Rouditchenko, Andrew and Vondrick, Carl and McDermott, Josh and Torralba, Antonio},
  booktitle={European conference on computer vision},
  pages={587--604},
  year={2018},
  organization={Springer}
}

@article{sato2025audio,
  title={Audio-visual localization based on spatial relative sound order},
  author={Sato, Tomoya and Sugano, Yusuke and Sato, Yoichi},
  journal={Machine Vision and Applications},
  volume={36},
  number={4},
  pages={90},
  year={2025},
  publisher={Springer}
}

@article{zhang2025scaling,
  title={Scaling up audio-synchronized visual animation: An efficient training paradigm},
  author={Zhang, Lin and Cai, Zefan and Zhou, Yufan and Mo, Shentong and Lin, Jinhong and Wu, Cheng-En and Wei, Yibing and Zhang, Yijing and Zhang, Ruiyi and Xiao, Wen and others},
  journal={arXiv preprint arXiv:2508.03955},
  year={2025}
}

@inproceedings{geng2023dense,
  title={Dense-localizing audio-visual events in untrimmed videos: A large-scale benchmark and baseline},
  author={Geng, Tiantian and Wang, Teng and Duan, Jinming and Cong, Runmin and Zheng, Feng},
  booktitle={2023 IEEE/CVF Conference on Computer Vision and Pattern Recognition (CVPR)},
  pages={22942--22951},
  year={2023},
  organization={IEEE}
}

@article{zhang2025visaudio,
  title={ViSAudio: End-to-End Video-Driven Binaural Spatial Audio Generation},
  author={Zhang, Mengchen and Chen, Qi and Wu, Tong and Liu, Zihan and Lin, Dahua},
  journal={arXiv preprint arXiv:2512.03036},
  year={2025}
}

@article{pian2026omnisonic,
  title={Omnisonic: Towards universal and holistic audio generation from video and text},
  author={Pian, Weiguo and Kushwaha, Saksham Singh and Chen, Zhimin and Deng, Shijian and Wang, Kai and Guo, Yunhui and Tian, Yapeng},
  journal={arXiv preprint arXiv:2604.04348},
  year={2026}
}

@article{zheng2026mmag,
  title={MMAG: A Multi-Control Mixed Audio Generation Benchmark},
  author={Zheng, Zihao and Xu, Xuenan and Mei, Jiahao and Li, Yixuan and Lv, Minghao and Wu, Wen and Zhang, Chao and Wu, Mengyue},
  journal={arXiv preprint arXiv:2608.06900},
  year={2026}
}

@inproceedings{kushwaha2025vintage,
  title={Vintage: Joint video and text conditioning for holistic audio generation},
  author={Kushwaha, Saksham Singh and Tian, Yapeng},
  booktitle={2025 IEEE/CVF Conference on Computer Vision and Pattern Recognition (CVPR)},
  pages={13529--13539},
  year={2025},
  organization={IEEE}
}

@inproceedings{wang2025t2a,
  title={T2a-feedback: Improving basic capabilities of text-to-audio generation via fine-grained ai feedback},
  author={Wang, Zehan and Lei, Ke and Zhu, Chen and Huang, Jiawei and Zhou, Sashuai and Liu, Luping and Cheng, Xize and Ji, Shengpeng and Ye, Zhenhui and Jin, Tao and others},
  booktitle={Proceedings of the 63rd Annual Meeting of the Association for Computational Linguistics (Volume 1: Long Papers)},
  pages={23535--23547},
  year={2025}
}

@article{wang2025audioeval,
  title={Audioeval: Automatic dual-perspective and multi-dimensional evaluation of text-to-audio-generation},
  author={Wang, Hui and Zhao, Jinghua and Cheng, Junyang and Liu, Cheng and Jia, Yuhang and Sun, Haoqin and Zhou, Jiaming and Qin, Yong},
  journal={arXiv preprint arXiv:2510.14570},
  year={2025}
}

@inproceedings{wang2026tta,
  title={Tta-bench: A comprehensive benchmark for evaluating text-to-audio models},
  author={Wang, Hui and Liu, Cheng and Chen, Junyang and Liu, Haoze and Jia, Yuhang and Zhao, Shiwan and Zhou, Jiaming and Sun, Haoqin and Bu, Hui and Qin, Yong},
  booktitle={Proceedings of the AAAI Conference on Artificial Intelligence},
  volume={40},
  number={39},
  pages={33512--33520},
  year={2026}
}

@article{li2026benchmarking,
  title={Benchmarking Single-Factor Physical Video-to-Audio Generation},
  author={Li, Tingle and Gururani, Siddharth and Shih, Kevin J and Bhatt, Gantavya and Lee, Sang-gil and Kong, Zhifeng and Goel, Arushi and Anumanchipalli, Gopala and Liu, Ming-Yu},
  journal={arXiv preprint arXiv:2605.30339},
  year={2026}
}

@inproceedings{heydari2025immersediffusion,
  title={Immersediffusion: A generative spatial audio latent diffusion model},
  author={Heydari, Mojtaba and Souden, Mehrez and Conejo, Bruno and Atkins, Joshua},
  booktitle={ICASSP 2025-2025 IEEE International Conference on Acoustics, Speech and Signal Processing (ICASSP)},
  pages={1--5},
  year={2025},
  organization={IEEE}
}

@article{helin2024soloaudio,
  author    = {Wang, Helin and Hai, Jiarui and Lu, Yen-Ju and Thakkar, Karan and Elhilali, Mounya and Dehak, Najim},
  title     = {SoloAudio: Target Sound Extraction with Language-oriented Audio Diffusion Transformer},
  journal   = {arXiv},
  year      = {2024},
}

@article{ma2024clapsep,
  title={CLAPSep: Leveraging Contrastive Pre-trained Models for Multi-Modal Query-Conditioned Target Sound Extraction},
  author={Ma, Hao and Peng, Zhiyuan and Li, Xu and Shao, Mingjie and Wu, Xixin and Liu, Ju},
  journal={arXiv preprint arXiv:2402.17455},
  year={2024}
}

@article{team2026gemma,
  title={Gemma 4 technical report},
  author={Team, Gemma and Abd, Sherif El and Aggarwal, Vaibhav and Algayres, Robin and Andreev, Alek and Bachem, Olivier and Ballantyne, Ian and Brick, Cormac and C{\u{a}}rbune, Victor and Casbon, Michelle and others},
  journal={arXiv preprint arXiv:2607.02770},
  year={2026}
}

@article{ghosh2026audio,
  title={Audio flamingo next: Next-generation open audio-language models for speech, sound, and music},
  author={Ghosh, Sreyan and Goel, Arushi and Jayakumar, Kaousheik and Koroshinadze, Lasha and Anand, Nishit and Kong, Zhifeng and Gururani, Siddharth and Lee, Sang-gil and Kim, Jaehyeon and Aljafari, Aya and others},
  journal={arXiv preprint arXiv:2604.10905},
  year={2026}
}

@misc{wang2025recrirmonauralblindroom,
      title={Blind Room Impulse Response Identification via Reverberant Speech Spectrum Reconstruction}, 
      author={Pengyu Wang and Xiaofei Li},
      year={2025},
      eprint={2509.15628},
      archivePrefix={arXiv},
      primaryClass={eess.AS},
      url={https://arxiv.org/abs/2509.15628}, 
}

@article{ratnam2003blind,
  title={Blind estimation of reverberation time},
  author={Ratnam, Rama and Jones, Douglas L and Wheeler, Bruce C and O’Brien Jr, William D and Lansing, Charissa R and Feng, Albert S},
  journal={The Journal of the Acoustical Society of America},
  volume={114},
  number={5},
  pages={2877--2892},
  year={2003},
  publisher={Acoustical Society of America}
}

@article{deshmukh2024pam,
  title={PAM: Prompting Audio-Language Models for Audio Quality Assessment},
  author={Soham Deshmukh and Dareen Alharthi and Benjamin Elizalde and Hannes Gamper and Mahmoud Al Ismail and Rita Singh and Bhiksha Raj and Huaming Wang},
  journal={arXiv preprint arXiv:2402.00282},
  year={2023}
}

@article{zheng2024bat,
  title={Bat: Learning to reason about spatial sounds with large language models},
  author={Zheng, Zhisheng and Peng, Puyuan and Ma, Ziyang and Chen, Xie and Choi, Eunsol and Harwath, David},
  journal={arXiv preprint arXiv:2402.01591},
  year={2024}
}

@inproceedings{cumlin24_interspeech,
  title     = {{DNSMOS Pro: A Reduced-Size DNN for Probabilistic MOS of Speech}},
  author    = {Fredrik Cumlin and Xinyu Liang and Victor Ungureanu and Chandan {K. A. Reddy} and Christian Schüldt and Saikat Chatterjee},
  year      = {2024},
  booktitle = {{Interspeech 2024}},
  pages     = {4818--4822},
  doi       = {10.21437/Interspeech.2024-478},
  issn      = {2958-1796},
}

@article{mittag2021nisqa,
  title={NISQA: A deep CNN-self-attention model for multidimensional speech quality prediction with crowdsourced datasets},
  author={Mittag, Gabriel and Naderi, Babak and Chehadi, Assmaa and M{\"o}ller, Sebastian},
  journal={arXiv preprint arXiv:2104.09494},
  year={2021}
}

@inproceedings{girdhar2023imagebind,
  title={ImageBind: One Embedding Space To Bind Them All},
  author={Girdhar, Rohit and El-Nouby, Alaaeldin and Liu, Zhuang
and Singh, Mannat and Alwala, Kalyan Vasudev and Joulin, Armand and Misra, Ishan},
  booktitle={CVPR},
  year={2023}
}

@article{jin2026sonoworld,
  title={SonoWorld: From One Image to a 3D Audio-Visual Scene},
  author={Jin, Derong and Chen, Xiyi and Lin, Ming C and Gao, Ruohan},
  journal={arXiv preprint arXiv:2603.28757},
  year={2026}
}

@inproceedings{scheibler2018pyroomacoustics,
  title={Pyroomacoustics: A python package for audio room simulation and array processing algorithms},
  author={Scheibler, Robin and Bezzam, Eric and Dokmani{\'c}, Ivan},
  booktitle={2018 IEEE international conference on acoustics, speech and signal processing (ICASSP)},
  pages={351--355},
  year={2018},
  organization={IEEE}
}

@inproceedings{dagli2025see,
  title={See-2-sound: Zero-shot spatial environment-to-spatial sound},
  author={Dagli, Rishit and Prakash, Shivesh and Wu, Robert and Khosravani, Houman},
  booktitle={Proceedings of the Special Interest Group on Computer Graphics and Interactive Techniques Conference Posters},
  year={2025}
}

@misc{yang2026controlfoleyunifiedcontrollablevideotoaudio,
  title={ControlFoley: Unified and Controllable Video-to-Audio Generation with Cross-Modal Conflict Handling}, 
  author={Jianxuan Yang and Xinyue Guo and Zhi Cheng and Kai Wang and Lipan Zhang and Jinjie Hu and Qiang Ji and Yihua Cao and Yihao Meng and Zhaoyue Cui and Mengmei Liu and Meng Meng and Jian Luan},
  year={2026},
  eprint={2604.15086},
  archivePrefix={arXiv},
  primaryClass={cs.MM},
  url={https://arxiv.org/abs/2604.15086}, 
}

@inproceedings{cheng2025mmaudio,
  title={Mmaudio: Taming multimodal joint training for high-quality video-to-audio synthesis},
  author={Cheng, Ho Kei and Ishii, Masato and Hayakawa, Akio and Shibuya, Takashi and Schwing, Alexander and Mitsufuji, Yuki},
  booktitle={2025 IEEE/CVF Conference on Computer Vision and Pattern Recognition (CVPR)},
  pages={28901--28911},
  year={2025},
  organization={IEEE}
}

@inproceedings{sun2025both,
  title={Both ears wide open: Towards language-driven spatial audio generation},
  author={Sun, Peiwen and Cheng, Sitong and Li, Xiangtai and Ye, Zhen and Liu, Huadai and Zhang, Honggang and Xue, Wei and Guo, Yike},
  booktitle={International Conference on Learning Representations},
  volume={2025},
  pages={42606--42649},
  year={2025}
}

@inproceedings{liu2026prismaudio,
  title={PrismAudio: Decomposed Chain-of-Thought and Multi-dimensional Rewards for Video-to-Audio Generation},
  author={Liu, Huadai and Luo, Kaicheng and Wang, Wen and Chen, Qian and Sun, Peiwen and Huang, Rongjie and Li, Xiangang and Ye, Jieping and Xue, Wei},
  booktitle={International Conference on Learning Representations},
  volume={2026},
  pages={25176--25204},
  year={2026}
}

@article{liu2026thinksound,
  title={ThinkSound: Chain-of-Thought Reasoning in Multimodal LLMs for Audio Generation and Editing},
  author={Liu, Huadai and Luo, Kaicheng and Wang, Jialei and Wang, Wen and Chen, Qian and Zhao, Zhou and Xue, Wei},
  journal={Advances in Neural Information Processing Systems},
  volume={38},
  pages={78526--78552},
  year={2026}
}

@inproceedings{tian2026audiox,
  title={AudioX: a unified framework for anything-to-audio generation},
  author={Tian, Zeyue and Liu, Zhaoyang and Jin, Yizhu and Yuan, Ruibin and Xue, Liumeng and Tan, Xu and Chen, Qifeng and Xue, Wei and Guo, Yike},
  booktitle={International Conference on Learning Representations},
  volume={2026},
  pages={110625--110651},
  year={2026}
}

@inproceedings{carion2026sam,
  title={Sam 3: Segment anything with concepts},
  author={Carion, Nicolas and Gustafson, Laura and Hu, Yuan-Ting and Debnath, Shoubhik and Hu, Ronghang and Suris Coll-Vinent, Didac and Ryali, Chaitanya and Alwala, Kalyan Vasudev and Khedr, Haitham and Huang, Andrew and others},
  booktitle={International conference on learning representations},
  volume={2026},
  pages={138846--138923},
  year={2026}
}

@inproceedings{chen2020vggsound,
  title={Vggsound: A large-scale audio-visual dataset},
  author={Chen, Honglie and Xie, Weidi and Vedaldi, Andrea and Zisserman, Andrew},
  booktitle={ICASSP 2020-2020 IEEE International Conference on Acoustics, Speech and Signal Processing (ICASSP)},
  pages={721--725},
  year={2020},
  organization={IEEE}
}

@article{ephrat2018looking,
  title={Looking to listen at the cocktail party: A speaker-independent audio-visual model for speech separation},
  author={Ephrat, Ariel and Mosseri, Inbar and Lang, Oran and Dekel, Tali and Wilson, Kevin and Hassidim, Avinatan and Freeman, William T and Rubinstein, Michael},
  journal={arXiv preprint arXiv:1804.03619},
  year={2018}
}

@inproceedings{wang2021curated,
  title={A curated dataset of urban scenes for audio-visual scene analysis},
  author={Wang, Shanshan and Mesaros, Annamaria and Heittola, Toni and Virtanen, Tuomas},
  booktitle={ICASSP 2021-2021 IEEE International Conference on Acoustics, Speech and Signal Processing (ICASSP)},
  pages={626--630},
  year={2021},
  organization={IEEE}
}

@inproceedings{zhou2022audio,
  title={Audio--visual segmentation},
  author={Zhou, Jinxing and Wang, Jianyuan and Zhang, Jiayi and Sun, Weixuan and Zhang, Jing and Birchfield, Stan and Guo, Dan and Kong, Lingpeng and Wang, Meng and Zhong, Yiran},
  booktitle={European Conference on Computer Vision},
  pages={386--403},
  year={2022},
  organization={Springer}
}

@article{morgado2018self,
  title={Self-supervised generation of spatial audio for 360 video},
  author={Morgado, Pedro and Vasconcelos, Nuno and Langlois, Timothy and Wang, Oliver},
  journal={Advances in neural information processing systems},
  volume={31},
  year={2018}
}

@article{cokelek2025spherical,
  title={Spherical Vision Transformers for Audio-Visual Saliency Prediction in {360\textdegree{}} Videos},
  author={Cokelek, Mert and Ozsoy, Halit and Imamoglu, Nevrez and Ozcinar, Cagri and Ayhan, Inci and Erdem, Erkut and Erdem, Aykut},
  journal={IEEE transactions on pattern analysis and machine intelligence},
  year={2025},
  publisher={IEEE}
}

@article{liu2025omniaudio,
  title={Omniaudio: Generating spatial audio from 360-degree video},
  author={Liu, Huadai and Luo, Tianyi and Luo, Kaicheng and Jiang, Qikai and Sun, Peiwen and Wang, Jialei and Huang, Rongjie and Chen, Qian and Wang, Wen and Li, Xiangtai and others},
  journal={arXiv preprint arXiv:2504.14906},
  year={2025}
}

@inproceedings{kim2025visage,
  title={Visage: Video-to-spatial audio generation},
  author={Kim, Jaeyeon and Yun, Heeseung and Kim, Gunhee},
  booktitle={International Conference on Learning Representations},
  volume={2025},
  pages={46404--46424},
  year={2025}
}

@article{tjandra2025meta,
  title={Meta audiobox aesthetics: Unified automatic quality assessment for speech, music, and sound},
  author={Tjandra, Andros and Wu, Yi-Chiao and Guo, Baishan and Hoffman, John and Ellis, Brian and Vyas, Apoorv and Shi, Bowen and Chen, Sanyuan and Le, Matt and Zacharov, Nick and others},
  journal={arXiv preprint arXiv:2502.05139},
  year={2025}
}

@article{wang2025berp,
  title={BERP: A blind estimator of room parameters for single-channel noisy speech signals},
  author={Wang, Lijun and Lu, Yixian and Gao, Ziyan and Li, Kai and Huang, Jianqiang and Kong, Yuntao and Okada, Shogo},
  journal={IEEE Transactions on Audio, Speech and Language Processing},
  volume={33},
  pages={2215--2231},
  year={2025},
  publisher={IEEE}
}

@article{xu2025qwen3,
  title={Qwen3-omni technical report},
  author={Xu, Jin and Guo, Zhifang and Hu, Hangrui and Chu, Yunfei and Wang, Xiong and He, Jinzheng and Wang, Yuxuan and Shi, Xian and He, Ting and Zhu, Xinfa and others},
  journal={arXiv preprint arXiv:2509.17765},
  year={2025}
}

@inproceedings{vyas2026pushing,
  title={Pushing the frontier of audiovisual perception with large-scale multimodal correspondence learning},
  author={Vyas, Apoorv and Chang, Heng-Jui and Yang, Cheng-Fu and Huang, Po-Yao and Gao, Luya and Richter, Julius and Chen, Sanyuan and Le, Matthew and Doll{\'a}r, Piotr and Feichtenhofer, Christoph and others},
  booktitle={Proceedings of the IEEE/CVF Conference on Computer Vision and Pattern Recognition},
  pages={30172--30182},
  year={2026}
}

@inproceedings{iashin2024synchformer,
  title={Synchformer: Efficient synchronization from sparse cues},
  author={Iashin, Vladimir and Xie, Weidi and Rahtu, Esa and Zisserman, Andrew},
  booktitle={ICASSP 2024-2024 IEEE International Conference on Acoustics, Speech and Signal Processing (ICASSP)},
  pages={5325--5329},
  year={2024},
  organization={IEEE}
}

@article{wilkins2023two,
  title={Two vs. four-channel sound event localization and detection},
  author={Wilkins, Julia and Fuentes, Magdalena and Bondi, Luca and Ghaffarzadegan, Shabnam and Abavisani, Ali and Bello, Juan Pablo},
  journal={arXiv preprint arXiv:2309.13343},
  year={2023}
}

@inproceedings{bochkovskiy2025depth,
  title={Depth pro: Sharp monocular metric depth in less than a second},
  author={Bochkovskiy, Alexey and Delaunoy, Ama{\"e}l and Germain, Hugo and Santos, Marcel and Zhou, Yichao and Richter, Stephan and Koltun, Vladlen},
  booktitle={International Conference on Learning Representations},
  volume={2025},
  pages={75602--75637},
  year={2025}
}

@article{shi2025sam,
  title={Sam audio: Segment anything in audio},
  author={Shi, Bowen and Tjandra, Andros and Hoffman, John and Wang, Helin and Wu, Yi-Chiao and Gao, Luya and Richter, Julius and Le, Matt and Vyas, Apoorv and Chen, Sanyuan and others},
  journal={arXiv preprint arXiv:2512.18099},
  year={2025}
}

@article{shimada2025open,
  title={Open-vocabulary sound event localization and detection with joint learning of clap embedding and activity-coupled cartesian doa vector},
  author={Shimada, Kazuki and Uchida, Kengo and Koyama, Yuichiro and Shibuya, Takashi and Takahashi, Shusuke and Mitsufuji, Yuki and Kawahara, Tatsuya},
  journal={IEEE Transactions on Audio, Speech and Language Processing},
  volume={33},
  pages={2946--2960},
  year={2025},
  publisher={IEEE}
}

@inproceedings{sun2025t2v,
  title={T2v-compbench: A comprehensive benchmark for compositional text-to-video generation},
  author={Sun, Kaiyue and Huang, Kaiyi and Liu, Xian and Wu, Yue and Xu, Zihan and Li, Zhenguo and Liu, Xihui},
  booktitle={2025 IEEE/CVF Conference on Computer Vision and Pattern Recognition (CVPR)},
  pages={8406--8416},
  year={2025},
  organization={IEEE}
}

@inproceedings{gao20192,
  title={{2.5D} visual sound},
  author={Gao, Ruohan and Grauman, Kristen},
  booktitle={2019 IEEE/CVF Conference on Computer Vision and Pattern Recognition (CVPR)},
  pages={324--333},
  year={2019},
  organization={IEEE}
}

@inproceedings{dixit2026foleybench,
  title={{FoleyBench}: A benchmark for video-to-audio models},
  author={Dixit, Satvik and Saito, Koichi and Zhong, Zhi and Mitsufuji, Yuki and Donahue, Chris},
  booktitle={ICASSP 2026-2026 IEEE International Conference on Acoustics, Speech and Signal Processing (ICASSP)},
  pages={14512--14516},
  year={2026},
  organization={IEEE}
}

@article{huang2023t2i,
  title={{T2I-CompBench}: A comprehensive benchmark for open-world compositional text-to-image generation},
  author={Huang, Kaiyi and Sun, Kaiyue and Xie, Enze and Li, Zhenguo and Liu, Xihui},
  journal={Advances in Neural Information Processing Systems},
  volume={36},
  pages={78723--78747},
  year={2023}
}

@inproceedings{schorkhuber2018binaural,
  title={Binaural rendering of ambisonic signals via magnitude least squares},
  author={Sch{\"o}rkhuber, Christian and Zaunschirm, Markus and H{\"o}ldrich, Robert},
  booktitle={Proceedings of the DAGA},
  volume={44},
  pages={339--342},
  year={2018}
}

@inproceedings{chen2023nvas,
    title = {Novel-view Acoustic Synthesis},
    author = {Changan Chen and Alexander Richard and Roman Shapovalov and Vamsi Krishna Ithapu and Natalia Neverova and Kristen Grauman and Andrea Vedaldi},
    year = {2023},
    booktitle = {CVPR},
    }

@article{zhang2026vidaudio,
  title={{VidAudio-Bench}: Benchmarking {V2A} and {VT2A} Generation across Four Audio Categories},
  author={Zhang, Qian and Cao, Yuqin and Gao, Yixuan and Min, Xiongkuo},
  journal={arXiv preprint arXiv:2604.10542},
  year={2026}
}

@inproceedings{huang2024vbench,
  title={{VBench}: Comprehensive benchmark suite for video generative models},
  author={Huang, Ziqi and He, Yinan and Yu, Jiashuo and Zhang, Fan and Si, Chenyang and Jiang, Yuming and Zhang, Yuanhan and Wu, Tianxing and Jin, Qingyang and Chanpaisit, Nattapol and others},
  booktitle={2024 IEEE/CVF Conference on Computer Vision and Pattern Recognition (CVPR)},
  pages={21807--21818},
  year={2024},
  organization={IEEE}
}

@inproceedings{SS2A,
  title={Gotta Hear Them All: Towards Sound Source Aware Audio Generation},
  author={Guo, Wei and Wang, Heng and Ma, Jianbo and Cai, Weidong},
  booktitle={AAAI},
  year={2026}
}

@article{viertola2026video,
  title={Video Object Segmentation-Aware Audio Generation},
  author={Viertola, Ilpo and Iashin, Vladimir and Rahtu, Esa},
  journal={International Journal of Computer Vision},
  volume={134},
  number={7},
  pages={326},
  year={2026},
  publisher={Springer}
}

@article{yoshida2026scene2sound,
  title={Scene2Sound: Auditory-Grounded Soundscape Generation for 3D Gaussian Worlds},
  author={Yoshida, Masaki and Togo, Ren and Ogawa, Takahiro and Haseyama, Miki},
  journal={arXiv preprint arXiv:2608.00463},
  year={2026}
}

@article{li2025sounding,
  title={Sounding that object: Interactive object-aware image to audio generation},
  author={Li, Tingle and Huang, Baihe and Zhuang, Xiaobin and Jia, Dongya and Chen, Jiawei and Wang, Yuping and Chen, Zhuo and Anumanchipalli, Gopala and Wang, Yuxuan},
  journal={arXiv preprint arXiv:2506.04214},
  year={2025}
}

@inproceedings{karchkhadze2026stereofoley,
  title={Stereofoley: Object-aware stereo audio generation from video},
  author={Karchkhadze, Tornike and Chen, Kuan-Lin and Heydari, Mojtaba and Henzel, Robert and Toso, Alessandro and Souden, Mehrez and Atkins, Joshua},
  booktitle={ICASSP 2026-2026 IEEE International Conference on Acoustics, Speech and Signal Processing (ICASSP)},
  pages={16027--16031},
  year={2026},
  organization={IEEE}
}
\bibliographystyle{iclr2027_conference}

\clearpage
\appendix
\section{Appendix}

\subsection{Additional Details on Evaluation Dimensions}
In this section, we provide additional details on the evaluation dimensions, metric formulations, and the selection of metric hyperparameters.

\noindent \textbf{Semantic Alignment.}
We evaluate whether the generated audio contains the sound source depicted in the input video. We adopt PE-AV as an open-vocabulary audio--text similarity model. For each generated audio clip, we compute its similarity to textual descriptions of all source classes in our predefined vocabulary. The complete source vocabulary is provided in Table~\ref{tab:source_vocab}. Classes are
ranked according to the similarity between their text embeddings and the generated audio embedding, and the top-$K$ predictions are retained for evaluation.

\begin{wraptable}{r}{0.47\linewidth}
\vspace{-8pt}
\centering
\small
\caption{Effect of text prompt formulation on PE-AV classification accuracy
on ESC-50.}
\label{tab:peav_prompts}
\setlength{\tabcolsep}{4pt}
\renewcommand{\arraystretch}{1.08}
\begin{tabular}{l r}
\toprule
\textbf{PE-AV Prompt} & \textbf{Top-1 Acc.} \\
\midrule
\texttt{\{class label\}}                         
    & 89.75 \\
I can hear \texttt{\{class label\}}              
    & 90.75 \\
This is an audio of \texttt{\{class label\}}     
    & 91.00 \\
This is \texttt{\{class label\}}                 
    & 82.50 \\
This is a sound of \texttt{\{class label\}}      
    & 90.50 \\
I hear \texttt{\{class label\}} in this audio    
    & \textbf{93.75} \\
\bottomrule
\end{tabular}
\vspace{-6pt}
\end{wraptable}

\noindent \textit{Prompt selection.}
Since audio--text similarity can be sensitive to prompt formulation~\citep{kushwaha2023multimodal}, we evaluate several prompt templates on ESC-50. As shown in Table~\ref{tab:peav_prompts}, the template
``I hear \texttt{\{class label\}} in this audio'' achieves the highest Top-1 accuracy, and we therefore use it for all semantic evaluations.

For a generated audio clip $a_i$, let $\mathcal{T}_K(a_i)$ denote its top-$K$ predicted source classes. Let $\mathcal{G}(y_i)$ contain the ground-truth source class $y_i$ together with its predefined equivalent classes. We define the correctness of each clip as
\begin{equation}
c_i =
\begin{cases}
1, & \mathcal{T}_K(a_i) \cap \mathcal{G}(y_i) \neq \emptyset, \\
0, & \text{otherwise}.
\end{cases}
\end{equation}
The semantic alignment score is then computed as the proportion of correctly matched clips:
\begin{equation}
S_{\mathrm{sem}}
=
\frac{1}{N}
\sum_{i=1}^{N} c_i.
\end{equation}
A higher score indicates that the generated audio more consistently preserves the sound source depicted in the input video.
\begin{wraptable}{r}{0.30\linewidth}
\vspace{-5pt}
\centering
\small
\caption{Effect of $K$ on semantic matching accuracy on a held-out validation set.}
\label{tab:topk_semantic}
\setlength{\tabcolsep}{7pt}
\renewcommand{\arraystretch}{1.08}
\begin{tabular}{cc}
\toprule
\textbf{Top-$K$} & \textbf{Accuracy} \\
\midrule
1  & 83.3\% \\
3  & 96.7\% \\
5  & 100.0\% \\
10 & 100.0\% \\
\bottomrule
\end{tabular}
\vspace{-6pt}
\end{wraptable}

\noindent \textit{Selection of $K$.}
We determine $K$ using a randomly sampled held-out set of 30 clips.
As shown in Table~\ref{tab:topk_semantic}, Top-5 retrieval correctly
recovers the target semantic class for all validation samples, while increasing $K$ to 10 provides no further improvement. We therefore set $K=5$ throughout our experiments.

Using Top-5 retrieval also provides robustness to the inherent ambiguity of audio semantics. Acoustically similar classes may receive comparable PE-AV similarity scores even when their visual identities are distinct. For example, sounds such as \emph{pig oinking} and \emph{snoring} may exhibit similar acoustic characteristics. Allowing a small set of highly ranked candidates,
together with predefined equivalent classes, avoids overly penalizing such ambiguous cases.

Finally, unlike a fixed closed-set audio classifier, this formulation supports open-vocabulary evaluation: new source classes can be incorporated simply by adding their textual descriptions to the vocabulary, without retraining the classifier. This makes the metric particularly suitable for evaluating diverse, in-the-wild video-to-audio generation.

\begin{figure}[t]
    \centering
    \includegraphics[width=0.95\linewidth]{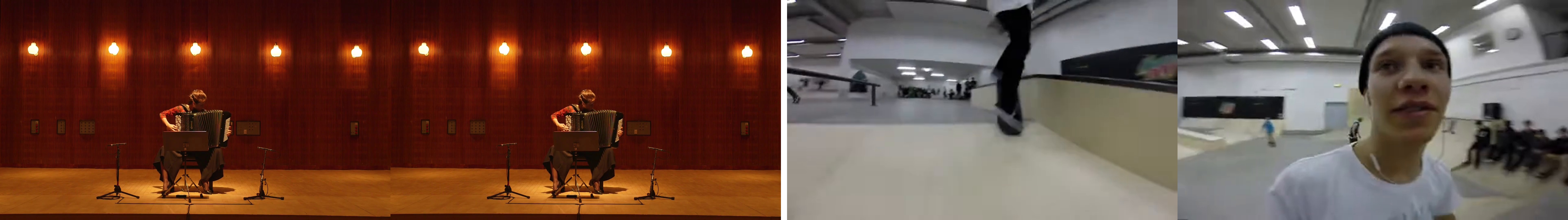}
    \caption{Example of single and multiple source semantics sound events.}
    \label{fig:semantics_eg}
\end{figure}

$\bullet$ \textbf{\textit{Sound Events}} -- \textbf{\textit{Multiple Sources}}.
For videos containing multiple sound sources, we evaluate each ground-truth source independently. For each source, we use its class label as a text prompt to extract the corresponding component from the generated audio using SAM-Audio. We then apply PE-AV to the extracted audio and retrieve its top-5 predicted classes. A source is considered correctly identified if its ground-truth class, or a predefined equivalent (e.g., tractor and bus), appears among these predictions.

For each video, we compute the fraction of correctly identified sources and then average this quantity across all videos:

\begin{equation}
S_{\mathrm{multi}}
=
\frac{1}{N}
\sum_{i=1}^{N}
\frac{\text{\# correctly identified sources in video } i}{M_i},
\end{equation}

where $M_i$ denotes the number of ground-truth sources in video $i$. This video-level averaging ensures that each video contributes equally to the final score, irrespective of the number of sources it contains.

Since SAM-Audio itself uses the source label as a text prompt, one potential concern is that source extraction could introduce a semantic bias. However, successful extraction alone is insufficient to obtain a correct score: the resulting audio must independently match the target class under PE-AV ranking over our large source vocabulary. Therefore, spurious agreement between the
extraction prompt and the subsequent semantic prediction is unlikely.

For SAM-Audio, we use FP16 precision for source extraction and FP32 precision for PE-AV re-ranking, with four re-ranking candidates. In our ablations, using FP32 for source extraction or increasing the number of re-ranking candidates from four to eight did not provide measurable improvements, so we use the more computationally efficient configuration throughout.

A qualitative example of the multi-source semantic evaluation is shown in Fig.~\ref{fig:semantics_eg}.

%%%%%%%%%%%%%%%%%%%%%%%%%%%%%%%%%%%%%%%%%%%%%%%%%%%%%
\begin{table*}[t]
\centering
\caption{Sound Source vocabulary used in our benchmark for semantic presence checking.}
\label{tab:source_vocab}

\noindent\rule{\textwidth}{0.5pt}
% \vspace{4pt}
\begin{multicols}{2}
\small
\begin{description}[
    leftmargin=0pt,
    labelindent=0pt,
    itemindent=0pt,
    itemsep=2pt,
    parsep=0pt
]

\item[\textbf{Transportation \& Vehicles.}]
motorcycle, motorboat, airplane, helicopter, train, tractor,
vehicle horn, snowmobile, car, car doors, car windows, car wiper,
bicycle, bus.

\item[\textbf{Alarms, Bells \& Signals.}]
police siren, ambulance siren, fire truck siren, alarm clock,
smoke detector, air horn, bell ringing, wind chime.

\item[\textbf{Animals, Birds \& Insects.}]
bird, goose, dog, cat, frog, lion, chicken, cattle, sheep, pig,
elephant, goat, whale, cow, zebra, cricket chirping, insect buzzing,
gibbon howling, horse clip-clop.

\item[\textbf{Human Vocalizations.}]
scream, booing, crowd, whistling, kid speaking, man speaking,
woman speaking, baby crying, laughing, coughing, singing choir,
child singing, female singing, male singing, baby babbling,
beatboxing, burping, nose blowing, slurping, sneezing, sobbing.

\item[\textbf{Human Actions.}]
clapping, eating, slapping, finger snapping, tap dancing,
keyboard typing, mouse clicking, writing sound, footsteps,
rope skipping.

\item[\textbf{Music, Instruments \& Ensembles.}]
singing bowl, guitar, bass guitar, double bass, cello, violin,
piano, drums, trumpet, trombone, saxophone, flute, clarinet,
french horn, cornet, harp, accordion, banjo, mandolin, ukulele,
sitar, erhu, theremin, organ, vibraphone, xylophone, glockenspiel,
harpsichord, harmonica, didgeridoo, bagpipes, percussion instrument,
synthesizer, tambourine, gong, oboe, steelpan, tuning fork, handpan,
ghatam, orchestra, bassoon, tabla, cajon, zither.

\item[\textbf{Sports \& Games.}]
basketball bounce, tennis, badminton, volleyball, skiing,
skateboarding, pool striking, football kicking, darts,
bowling sounds, golf swing.

\item[\textbf{Household Appliances.}]
vacuum cleaner, electric blender, toilet flushing, electric shaver,
washing machine, electric fan, air conditioning noise, hair dryer.

\item[\textbf{Machines \& Tools.}]
chainsaw, printer, lawn mowing, lathe spinning, hammer sound,
chopping sound, arc welding.

\item[\textbf{Household Object Interactions.}]
liquid pouring, opening drawers, closing drawers, scraping,
door sound, object dragging, knife sharpening.

\item[\textbf{Nature, Weather \& Water.}]
volcano explosion, water flowing, rain, underwater bubbling,
thunder, sea waves, wind blowing, waterfall.

\item[\textbf{Impacts, Weapons \& Fire.}]
fireworks, missile, fire crackling, gun.

\end{description}
\end{multicols}

\noindent\rule{\textwidth}{0.5pt}
\end{table*}
%%%%%%%%%%%%%%%%%%%%%%%%%%%%%%%%%%%%%%%%%%%%%%%%%%%%%%

\noindent \textbf{Soundscape.}
We evaluate whether the generated audio is consistent with the acoustic environment depicted in the input video using an audio-only scene classifier based on Qwen3-Omni-30B-A3B-Thinking. Our evaluation set contains 100 clips from the TAU Urban Acoustic Scenes dataset, balanced across 10 classes with 10 clips per class: \emph{airport}, \emph{bus}, \emph{metro}, \emph{metro station}, \emph{park}, \emph{public square}, \emph{shopping mall}, \emph{pedestrian street}, \emph{street traffic}, and \emph{tram}.

For each generated soundtrack, \textit{Qwen3-Omni-30B-A3B-Thinking} analyzes acoustic cues such as enclosure, listener motion, road or rail activity, crowd presence, natural ambience, traffic, and announcements, and ranks the five most likely scene classes. We report Top-3 accuracy, counting a sample as correct if its ground-truth scene appears among the three highest-ranked predictions. Below is the prompt used.

We use Top-3 rather than Top-1 because several urban soundscapes are inherently ambiguous from short audio alone. For example, \emph{metro}, \emph{metro station}, and \emph{tram} may share rail and announcement cues, while \emph{airport}, \emph{shopping mall}, and \emph{metro station} can exhibit similar crowd and reverberation characteristics. Likewise, \emph{public square}, \emph{pedestrian street}, and \emph{street traffic} may contain overlapping traffic and crowd sounds. Requiring exact Top-1 classification can therefore penalize acoustically plausible generations due to evaluator uncertainty rather than a genuine soundscape mismatch. Top-3 provides limited tolerance to this ambiguity while remaining sufficiently selective.

\begin{small}
\begin{verbatim}
Analyze this recording as a closed-set acoustic scene.

First determine only these acoustic attributes. Use "uncertain" whenever direct
evidence is absent:

- enclosure: indoor, outdoor, or uncertain
- listener_motion: moving, stationary, or uncertain
- road_vehicle_evidence: present, absent, or uncertain
- rail_evidence: present, absent, or uncertain
- crowd_activity: low, medium, high, or uncertain
- nature_activity: low, medium, high, or uncertain
- traffic_activity: low, medium, high, or uncertain
- announcement_evidence: present, absent, or uncertain

Important rules:
- Do not infer a source or location from timbre alone.
- Generic rumble, hum, reverberation, or metallic noise is not rail evidence.
- Rail evidence requires a consistent combination of unmistakable track or
  wheel behavior and motion or station context.
- Do not infer motion unless the audio changes consistently over time.
- Do not invent cues that are not clearly audible.
- Treat all classes as equally likely before listening.

After extracting the attributes, evaluate every candidate independently.
A compatible but non-specific cue provides weak evidence. Contradictory
attributes should strongly lower a candidate.

Rank five distinct labels:
[airport, bus, metro, metro_station, park, public_square, shopping_mall,
 street_pedestrian, street_traffic, tram]

Return only:
{"class_pred":["rank1","rank2","rank3","rank4","rank5"],
 "reason":"Explain rank 1 using only the extracted attributes and explicitly state 
 why the closest alternative ranked lower."}
\end{verbatim}
\end{small}

%%%%%%%%%%%%%%%%%%%%%%%%%%%%%%%%%%%%%%%%

\noindent \textbf{Temporal Alignment.}
We evaluate whether the generated audio contains the correct sound source and whether that source is synchronized with its visual activity. A sample receives temporal credit only when both conditions are satisfied.

We evaluate three settings: \textit{sustained}, \textit{transient}, and \textit{multi-source}. Sustained events remain active over longer periods, whereas transient events correspond to short, localized sound events. Multi-source videos are evaluated independently for each annotated source.

\noindent \textit{Semantic verification.}
We first verify whether the target sound is present using PE-AV. The generated audio is compared against our source vocabulary using the prompt ``I hear \texttt{\{class\}} in this audio.'' A source is considered present if its ground-truth class, or an accepted equivalent, appears among the top-5 predictions.

\noindent \textit{Temporal synchronization.}
For each target source, we use SAM-Audio to extract the corresponding sound from the generated audio. The visual input is masked to retain only the target source, reducing interference from unrelated motion. We then use DeSync to estimate the audio--visual temporal offset. A source is considered synchronized when
\begin{equation}
d_i \leq 0.7 \text{ s},
\end{equation}
where $d_i$ denotes the predicted absolute audio--visual offset. Lower DeSync values indicate better temporal alignment.
\noindent \textit{Single-source evaluation.}
For single-source videos, let $s_i$ indicate whether the target sound is semantically present and let $t_i$ indicate whether it satisfies the temporal threshold. We report the joint semantic--temporal score
\begin{equation}
S_{\mathrm{temp}}
=
\frac{1}{N}
\sum_{i=1}^{N}
s_i t_i.
\end{equation}
Thus, a sample receives credit only when the correct sound is generated and is temporally synchronized with the corresponding visual event. Sustained and transient subsets use the same metric and are reported separately.

\noindent \textit{Multi-source evaluation.}
For videos containing multiple sources, we first extract each annotated source independently using SAM-Audio. PE-AV then verifies the semantics of each extracted source, and DeSync measures its synchronization with the corresponding masked visual source.
For video $i$ with $M_i$ annotated sources, we compute
\begin{equation}
S_{\mathrm{temp}}^{(i)}
=
\frac{1}{M_i}
\sum_{j=1}^{M_i}
s_{ij} t_{ij},
\end{equation}
and average this score across videos:
\begin{equation}
S_{\mathrm{temp}}^{\mathrm{multi}}
=
\frac{1}{N}
\sum_{i=1}^{N}
S_{\mathrm{temp}}^{(i)}.
\end{equation}
This source-wise formulation ensures that the correct sound must be synchronized with the correct visual source, while the per-video averaging prevents videos with more annotated sources from receiving greater weight.

% \noindent \textbf{Temporal Alignment.} This dimension evaluates the source-aware temporal alignment between video and its corresponding audio.
% \comment{Maybe we will divide the classes based on how they react to sensitive changes in the time shifts. This will answer the question about how transient and sustained are selected.}

% \begin{itemize}[leftmargin=1.2em, labelsep=0.5em, topsep=0pt, itemsep=1pt]
% \item \textbf{Single Source} -- \textbf{Sustained.} We first verify that the generated audio contains the target sustained sound and isolate it using SAM-Audio. We then use Synchformer~\citep{iashin2024synchformer} to measure its temporal synchronization with the corresponding visual activity \takashi{Koichi asks whether this is the same as DeSync and whether we mask. We should be precise: state exactly what Synchformer output we use (e.g., DeSync score vs. offset), and whether we mask the video/audio to the target source before feeding Synchformer. This directly affects reproducibility, so at least one clarifying sentence in the main text is worth the space.}.

% \item \textbf{Single Source} -- \textbf{Transient.} We isolate the target transient sound and use Synchformer to measure whether its onset and timing align with the corresponding visual event.

% \item \textbf{Multiple Sources.} We identify and isolate each sound source using SAM-Audio, then use Synchformer to evaluate the temporal alignment of each source with its corresponding visual event.
% \end{itemize}
%%%%%%%%%%%%%%%%%%%%%%%%%%%%%%%%%%%%%%%%

\noindent \textbf{Spatial Alignment -- Single Source.}
We evaluate whether the generated sound is spatialized at the same
\emph{left}, \emph{center}, or \emph{right} location as the visible source. For static clips, we use the annotated source location directly. For moving clips, we divide the video into one-second windows and compute the median horizontal position of the source mask within each window. The normalized horizontal position is quantized into left, center, and right regions using
boundaries of $0.45$ and $0.55$.

\noindent \textit{Audio localization.}
For stereo audio, we estimate the source direction using the normalized left--right energy difference
\begin{equation}
c =
\frac{E_L-E_R}{E_L+E_R},
\qquad
E_L=\sum_t L_t^2,\quad
E_R=\sum_t R_t^2.
\end{equation}
We compare the observed cue $c$ with the expected cues produced by our stereo renderer for candidate azimuths between $-90^\circ$ and $90^\circ$, sampled at $2^\circ$ intervals, and select the closest matching direction. For binaural audio, we instead estimate direction by matching interaural level and phase differences to HRIR-based directional templates.

The estimated azimuth $\hat{\theta}$ is projected onto the horizontal video coordinate using the camera field of view $\alpha$:
\begin{equation}
\hat{x}
=
\frac{1}{2}
-
\frac{\tan\hat{\theta}}
     {2\tan(\alpha/2)}.
\end{equation}
We then apply the same $0.45/0.55$ boundaries to obtain the predicted left/center/right audio location.

\noindent \textit{Source-aware spatial score.}
For clip $i$, let $r_i$ denote the fraction of evaluated windows for which the predicted audio location agrees with the visual source location. For static clips, $r_i$ corresponds to a single left/center/right comparison; for moving clips, it is averaged over all valid one-second windows.

We additionally use PE-AV to verify that the generated audio contains the target source. Let $s_i=1$ if the ground-truth class, or an accepted equivalent, appears in the PE-AV top-5 predictions, and $s_i=0$ otherwise. We report
\begin{equation}
S_{\mathrm{loc}}
=
\frac{\sum_i s_i r_i}
     {\sum_i s_i},
\qquad
S_{\mathrm{SemLoc}}
=
\frac{1}{N}\sum_{i=1}^{N}s_i r_i.
\end{equation}
Here, $S_{\mathrm{loc}}$ measures localization accuracy only among clips in which the correct sound source is present, while $S_{\mathrm{SemLoc}}$ is the stricter source-aware metric: a clip receives credit only when the correct sound is generated \emph{and} spatialized at the correct visual location.

%%%%%%%%%%%%%%%%%%%%%%%%%%%%%%
% $ \bullet$ \textbf{\textit{Multiple Sources}}. We first apply source gating to identify the sound sources present in the generated audio \takashi{Agree with Koichi. ``We first apply source gating to identify the sound sources present'' repeats what we already stated in the source-gating paragraph. We can safely delete this sentence and start directly with ``We localize each detected source independently...''. This also buys us a little space.} and then localize each detected source independently. The estimated audio locations are matched with their corresponding visual sources to measure source-wise spatial alignment \takashi{Koichi wants the matching ``how''. Example: ``matched to visual sources via Hungarian assignment on semantic labels'' (or whatever we actually use). If it's simple label matching, say so explicitly. Details $\to$ appendix.}.\\
\noindent \textbf{Spatial Alignment -- Multiple Sources.}
For videos containing multiple sound sources, we evaluate whether the generated audio preserves the \emph{relative left--right ordering} of the visible sources. Each ground-truth source class is first used as a text prompt to extract the corresponding component from the generated two-channel audio using SAM-Audio.
We then verify the semantics of each extracted source with PE-AV. Source $j$ receives a semantic hit, $s_j=1$, if its ground-truth class, or an accepted equivalent, appears among the top-5 predictions.

\noindent \textit{Visual and audio positions.}
For each source, the visual horizontal position $x_j^{V}$ is obtained by computing the median horizontal position of its mask in each frame and then taking the median across the full clip. To estimate its audio position, we use interaural cues from the extracted stereo signal: low-frequency interaural time
differences (ITD; 50--620\,Hz) and high-frequency interaural level differences (ILD; 1.7--4.6\,kHz). These cues are combined to obtain a continuous horizontal audio position $x_j^{A}$.

Unlike the single-source metric, which uses left/center/right regions, the multi-source metric directly compares these continuous source positions over the entire clip.
For each pair of sources $(a,b)$, we define the relative spatial correctness as
\begin{equation}
r_{ab}
=
\mathbf{1}\!\left[
(x_b^{V}-x_a^{V})(x_b^{A}-x_a^{A}) > 0
\right].
\end{equation}
Thus, a pair is correct when the source that appears farther to the right in the video is also localized farther to the right in the generated audio.

Both sources must also be semantically identified for the pair to receive source-aware credit. We therefore define
\begin{equation}
s_{ab}=s_a s_b.
\end{equation}
Across all $P$ source pairs, we report
\begin{equation}
S_{\mathrm{rel}}
=
\frac{1}{P}
\sum_{(a,b)} r_{ab},
\end{equation}
\begin{equation}
S_{\mathrm{cond}}
=
\frac{\sum_{(a,b)} s_{ab}r_{ab}}
     {\sum_{(a,b)} s_{ab}},
\qquad
S_{\mathrm{joint}}
=
\frac{1}{P}
\sum_{(a,b)} s_{ab}r_{ab}.
\end{equation}
Here, $S_{\mathrm{rel}}$ measures relative spatial ordering alone,
$S_{\mathrm{cond}}$ measures ordering among semantically valid source pairs,
and $S_{\mathrm{joint}}$ is the stricter source-aware metric, requiring both
sources to be correctly identified and correctly ordered in space.
In particular, $S_{\mathrm{cond}}$ isolates localization ability from semantic detection. Because it conditions on correct source identification, a drop in $S_{\mathrm{cond}}$ reflects spatial-ordering errors rather than semantic failures.

\noindent \textbf{Depth-loudness consistency.}
We evaluate whether the generated sound becomes louder as the visible source approaches the camera and quieter as it moves away. For each video, Depth Pro estimates the source depth in every frame. We compute the median depth within the annotated source mask and aggregate these estimates into one-second values $d_{i,t}$. In parallel, SAM-Audio extracts the target sound from the generated audio, and its RMS amplitude within each one-second window gives $e_{i,t}$.

A transition is considered correct when depth and audio intensity change in opposite directions:
\begin{equation}
h_{i,t}
=
\mathbf{1}\!\left[
(d_{i,t+1}-d_{i,t})
(e_{i,t+1}-e_{i,t}) < 0
\right].
\end{equation}
Thus, a source moving closer while becoming louder, or moving farther while becoming quieter, receives credit. Transitions with no change, or without a valid depth estimate, are not counted as correct.

We additionally verify that the generated clip contains the target sound using PE-AV. Let $s_i=1$ if the ground-truth class, or an accepted equivalent, appears among the top-10 predictions, and $s_i=0$ otherwise. Let $T_i$ denote the number of valid transitions in video $i$.

For semantically correct clips, we first compute the mean distance-consistency score
\begin{equation}
S_{\mathrm{dist}}
=
\frac{\sum_{i:s_i=1}
\left(
\frac{1}{T_i}\sum_t h_{i,t}
\right)}
{\sum_i s_i}.
\end{equation}
We also report a joint semantic--distance score:
\begin{equation}
S_{\mathrm{SemDist}}
=
\frac{\sum_i s_i \sum_t h_{i,t}}
     {\sum_i T_i}.
\end{equation}
Here, $S_{\mathrm{dist}}$ measures depth--loudness consistency only among clips where the target sound is semantically present, giving each such video equal weight. In contrast, $S_{\mathrm{SemDist}}$ gives credit only when both the correct sound is present and the corresponding depth--loudness transition is consistent; since it is normalized over all valid transitions, longer clips may
contribute more.

\noindent \textbf{Acoustics.}
We evaluate whether the generated audio exhibits a reverberation level that is consistent with the visible environment. We use BERP to estimate the reverberation time, $\hat{T}_{60}$, i.e., the time required for the sound level to decay by 60\,dB. The estimated value is mapped to one of three categories:
\begin{equation}
\hat{c}_i =
\begin{cases}
\text{dry}, & \hat{T}_{60,i} < 0.3\,\text{s},\\
\text{high reverb}, & \hat{T}_{60,i} > 0.5\,\text{s},\\
\text{uncertain}, & \text{otherwise}.
\end{cases}
\end{equation}
A clip passes the reverberation check when the predicted category matches its annotated acoustic condition. Predictions in the intermediate $0.3$--$0.5$\,s range are treated as uncertain and receive no credit for either the dry or high-reverberation class.

We additionally verify that the generated audio contains the intended sound using PE-AV. Let $r_i=1$ if the predicted reverberation class is correct, and let $s_i=1$ if the ground-truth sound class, or an accepted equivalent, appears among the top-5 PE-AV predictions. Over $N$ clips, we report
\begin{equation}
S_{\mathrm{rev}}
=
\frac{1}{N}\sum_{i=1}^{N} r_i,
\qquad
S_{\mathrm{sem}}
=
\frac{1}{N}\sum_{i=1}^{N} s_i,
\qquad
S_{\mathrm{SemRev}}
=
\frac{1}{N}\sum_{i=1}^{N} r_i s_i.
\end{equation}
Here, $S_{\mathrm{rev}}$ measures reverberation consistency alone, while $S_{\mathrm{SemRev}}$ is the source-aware joint score, giving credit only when the generated audio contains the intended sound and exhibits the expected reverberation level.

\subsection{Benchmark statistics}
\begin{figure}[t]
    \centering
    \includegraphics[width=0.97\linewidth]{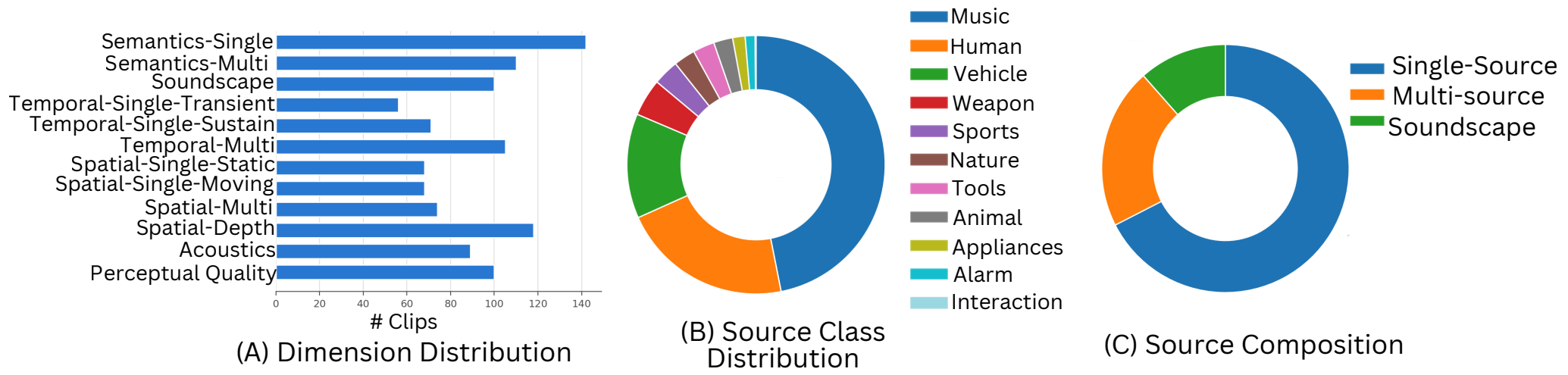}
    \caption{Benchmark stats. (A) Number of clips per evaluation dimension, (B) distribution over source classes, and (C) composition of single-source, multi-source, and soundscape clips.
    % \kazuki{We can re-order this into (A) Dimension Distribution, (B) Source Class Distribution, and (C) Source Type Distribution, since Dimension Distribution and Source Class Distribution are more informative. Additionally, we can rename ``Source Type Distribution'' into ``Source Composition'' as ``Type'' is ambiguous.}
    % \kazuki{Dimension Distribution's y-axis order: sustained first come before transient.}
    % \kazuki{Move to Appendix.}
    % \kazuki{Keep.}
    }
    \label{fig:stats}
\end{figure}

% As shown in Fig.~\ref{fig:stats}, the sources span a wide taxonomy including music, human and vehicles, weapon, sports, and household or environmental sounds.
% Figure~\ref{fig:stats}~(C) shows that the distributions across evaluation dimensions provide coverage over semantic, spatial, temporal, acoustic, and perceptual categories each category ranging from 50-150.
% A portion of the benchmark contains multiple simultaneously active sound sources, enabling explicit evaluation of source-aware generation. We have finally 767 uniques videos.
% As illustrated in Fig.~\ref{fig:metrics}~(D), the spatial subset spans a wide range of source directions: sources in the 360° $\rightarrow$ FOA setting are distributed across the full sphere, while those in the PoV $\rightarrow$ stereo/binaural setting spread broadly from left through center to right.
% The distance subset likewise spans varied source-to-camera distances. We also balance clips across reverberation conditions and diverse soundscapes, ensuring evaluation of semantic scene correctness and acoustic realism.
Figure~\ref{fig:stats}~(A) shows that the benchmark provides broad coverage across semantic, temporal, spatial, acoustic, and perceptual evaluation dimensions, with approximately 50--150 videos per dimension. As some videos are annotated for and reused across multiple dimensions, these per-dimension counts overlap, yielding 767 unique perspective videos overall.
Fig.~\ref{fig:stats}~(B) shows that, over the 767 unique videos, the benchmark spans a broad source taxonomy, including music, human sounds, vehicles, impulsive events, sports, tools, alarms, household sounds, and environmental sources. Music constitutes the largest source category, followed by human and vehicle-related sounds.
Fig.~\ref{fig:stats}~(C) shows that a substantial subset contains multiple simultaneously active sources, typically two to three per video, enabling explicit source-aware evaluation.
% \kazuki{I re-ordered the paragraph to Dimension Distribution, Source Class Distribution, and Source Type Distribution, in order to mention first the total video size. (Original: Source Class Distribution, Dimension Distribution, and Source Type Distribution). If we use this order, please also align the Fig.~\ref{fig:stats} to state like (A), (B), and (C) in the main sentences.}
% \kazuki{One question: Are Source Class Distribution and Source Composition computed on 767 unique videos? Or are some videos are counted twice or more like Dimension Distribution?}

\begin{table}[t]
\centering
\caption{Training data and model scales of the evaluated baseline methods, grouped by task. Since SEE-2-SOUND~\citep{dagli2025see} uses CoDi~\citep{tang2023any} in a zero-shot manner, we report CoDi instead of SEE-2-SOUND. Note that CoDi and AudioX are also trained on tasks other than audio generation; here we report only the training data relevant to audio generation. N/R denotes not reported.}
\label{tab:model_scale}
\setlength{\tabcolsep}{6pt}
\renewcommand{\arraystretch}{1.1}
\resizebox{\textwidth}{!}{%
\begin{tabular}{lccc}
\toprule
\textbf{Method} & \textbf{\# of Train Samples} & \textbf{Train Data Hours} & \textbf{Trainable Params} \\
\midrule
\multicolumn{4}{l}{\textit{Mono Audio Generation}} \\
CoDi~\citep{tang2023any} & 4.48M & N/R & N/R \\
MMAudio~\citep{cheng2025mmaudio} & 1.13M & 2.61k & 1.03B \\
\midrule
\multicolumn{4}{l}{\textit{Stereo Audio Generation}} \\
AudioX~\citep{tian2026audiox} & 1.69M & 4.99k & 1.1B \\
ThinkSound~\citep{liu2026thinksound} & N/R & 2.53k & 1.3B \\
PrismAudio~\citep{liu2026prismaudio} & N/R & 8.13k & 518M \\
SpatialSonic~\citep{sun2025both} & 1.02M & 2.85k & N/R \\
\midrule
\multicolumn{4}{l}{\textit{FOA Audio Generation}} \\
ViSAGe~\citep{kim2025visage} & 0.28M & 0.67k & 360M \\
\midrule
\multicolumn{4}{l}{\textit{Audio Spatialization}} \\
Mono2Binaural~\citep{gao20192} & 1.50K & 4.16 & N/R \\
\bottomrule
\end{tabular}
}
\end{table}

\subsection{Additional Experiments}

\begin{figure}[t]
    \centering
    \includegraphics[width=0.97\linewidth]{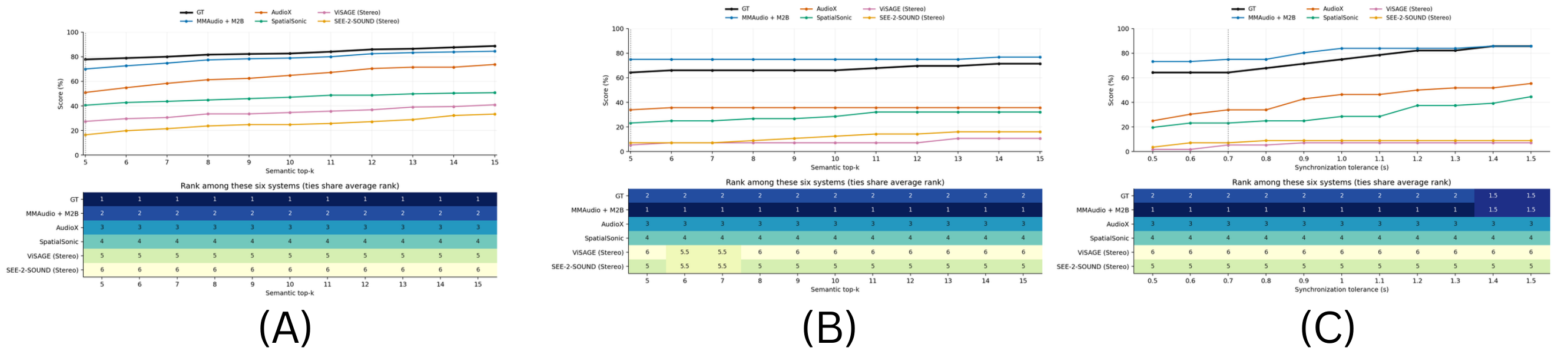}
    \caption{Metric robustness. Increasing top-$k$ from 5 to 15 improves (A) GT multi-source semantics from 77.7\% to 88.7\% and, due to semantic gating, (B) transient synchronization from 64.2\% to 71.4\%.
    Similarly, relaxing the synchronization tolerance from 0.5 to 1.5 s further increases (C) GT transient synchronization from 64.2\% to 85.7\% (keeping $k=5$ fixed). Importantly, baseline rankings remain consistent across these settings, supporting the robustness of our relative comparisons.}
    \label{fig:gt_ceil}
\end{figure}

\noindent \paragraph{Decomposing evaluation scores.}
To better understand evaluator errors, we decompose the GT scores into source-presence accuracy, evaluator accuracy conditioned on semantic correctness, and the final joint score (Table~\ref{tab:score_decomposition}).

\begin{table}[h]
\centering
\caption{Decomposition of ground-truth evaluation scores into
source-presence accuracy, conditional evaluator scores, and
final joint scores. All values are percentages except perceptual
quality (PQ).}
\label{tab:score_decomposition}
\scriptsize
\renewcommand{\arraystretch}{1.1}

\begin{tabular}{@{}lccc|ccc|cccc|cc@{}}
\toprule
& \multicolumn{3}{c|}{Semantics}
& \multicolumn{3}{c|}{Temporal}
& \multicolumn{4}{c|}{Spatial}
& Acoustics & PQ \\
\cmidrule(lr){2-4}
\cmidrule(lr){5-7}
\cmidrule(lr){8-11}
\cmidrule(lr){12-12}
\cmidrule(l){13-13}
Metric
& Single & Multi & Scene
& Sust. & Trans. & Multi
& Static & Moving & Multi & Depth
& Reverb & Quality \\
\midrule
Source presence
& 99.0 & 77.7 & 78.0
& 97.0 & 89.2 & 86.5
& 97.0 & 97.0 & 94.1 & 88.1
& 95.5 & -- \\

Evaluator score
& -- & -- & --
& 72.4 & 72.0 & 74.2
& 79.5 & 72.3 & 77.6 & 73.6
& 100.0 & 6.3 \\

\midrule
\textbf{Joint score}
& 99.0 & 77.7 & 78.0
& 70.4 & 64.2 & 65.2
& 77.2 & 70.2 & 68.6 & 72.8
& 95.5 & 6.3 \\
\bottomrule
\end{tabular}
\end{table}

\paragraph{Spatial cue preservation after separation.}
To assess whether source separation preserves spatial cues, we randomly sample 100 ground-truth stereo clips and compare localization accuracy before and after applying SAM-Audio independently to both channels (shown in Table~\ref{tab:spatial_separation}). These results suggest that source separation largely preserves the inter-channel spatial cues needed for localization.

\begin{table}[h]
\centering
\caption{Spatial localization accuracy (\%) before and after source separation on 100 GT stereo clips.}
\label{tab:spatial_separation}
\footnotesize
\setlength{\tabcolsep}{3pt}
\renewcommand{\arraystretch}{1.1}
\begin{tabular}{@{}cccc@{}}
\toprule
\shortstack{Before\\separation} &
\shortstack{After\\separation} &
\shortstack{Always\\center} &
Random \\
\midrule
\textbf{75.1} & 68.7 & 16.0 & 33.3 \\
\bottomrule
\end{tabular}
\end{table}

\paragraph{Decomposing multi-source spatial scores.}
Our multi-source spatial metric jointly evaluates source presence and localization, rather than localization alone. To distinguish these factors, we report pair-semantic accuracy, localization accuracy conditioned on semantic success, and their joint score (Table~\ref{tab:multi_spatial_decomp}). The joint metric also correlates strongly with human judgments ($r=0.93$, Sec.~4.3).

\begin{table}[h]
\centering
\caption{Decomposition of multi-source spatial scores (\%).
Localization accuracy is conditioned on both sources
passing semantic verification.}
\label{tab:multi_spatial_decomp}
\footnotesize
\setlength{\tabcolsep}{3pt}
\renewcommand{\arraystretch}{1.05}

\resizebox{0.8\columnwidth}{!}{%
\begin{tabular}{@{}lrrr@{}}
\toprule
Method &
\shortstack{Pair semantics\\(\%)} &
\shortstack{Localization\\(sem. passed, \%)} &
\shortstack{Joint\\(\%)} \\
\midrule
AudioX                  & 24.42 & \textbf{66.67} & \textbf{16.28} \\
ThinkSound              & \textbf{33.72} & 44.83 & 15.12 \\
PrismAudio              & 15.12 & 61.54 & 9.30 \\
SpatialSonic            & 10.47 & 44.44 & 4.65 \\
SEE-2-SOUND (Stereo)    & 5.81  & 60.00 & 3.49 \\
SEE-2-SOUND (Binaural)  & 2.33  & 0.00  & 0.00 \\
ViSAGe (Stereo)         & 13.95 & 25.00 & 3.49 \\
ViSAGe (Binaural)       & 19.77 & 52.94 & 10.47 \\
MMAudio + M2B           & \textbf{33.72} & 44.83 & 15.12 \\
\midrule
GT                      & 88.37 & 77.63 & 68.60 \\
\bottomrule
\end{tabular}%
}
\end{table}

\end{document}